\documentclass[desactivate]{aa}  
\usepackage{graphicx}
\usepackage{txfonts}
\usepackage{hyperref}

\newcommand{\rewrite}[1]{#1}
\newcommand{\drafttwo}[1]{#1}
\newcommand{\paraphrased}[1]{#1}
\newcommand{\referee}[1]{#1}

\newcommand{\changed}[1]{#1}

\begin{document}

   \title{Improving the open cluster census.}

   \subtitle{III. Using cluster masses, radii, and dynamics to create a cleaned open cluster catalogue\thanks{Tables~\ref{tab:catalogue_clusters}~and~\ref{tab:catalogue_members} are only available in electronic form at the CDS via anonymous ftp to cdsarc.cds.unistra.fr (130.79.128.5) or via https://cdsarc.cds.unistra.fr/cgi-bin/qcat?J/A+A/}
   }


   \author{Emily L. Hunt\inst{1} \and Sabine Reffert\inst{1}}

   \institute{Landessternwarte, Zentrum für Astronomie der Universität Heidelberg, Königstuhl 12, 69117 Heidelberg, Germany\\
              \email{ehunt@lsw.uni-heidelberg.de}
             }

   \date{Received 17 November 2023; accepted 5 March 2024}

 
\abstract
{The census of open clusters has exploded in size thanks to data from the Gaia satellite. However, it is likely that many of these reported clusters are not gravitationally bound, making the open cluster census impractical for many scientific applications.}
{We aim to test different physically motivated methods for distinguishing between bound and unbound clusters, using them to create a cleaned star cluster catalogue.}
{We derived completeness-corrected photometric masses for 6956 clusters from our earlier work. Then, we used these masses to compute the size of the Roche surface of these clusters (their Jacobi radius) and distinguish between bound and unbound clusters.}
{We find that only 5647 (79\%) of the clusters from our previous catalogue are compatible with bound open clusters, dropping to just 11\% of clusters within 250~pc. Our catalogue contains 3530 open clusters in a more strongly cut high-quality sample of objects. The moving groups in our sample show different trends in their size as a function of age and mass, suggesting that they are unbound and undergoing different dynamical processes. Our cluster mass measurements constitute the largest catalogue of Milky Way cluster masses to date, which we also use for further science. Firstly, we inferred the mass-dependent completeness limit of the open cluster census, showing that the census is complete within 1.8~kpc only for objects heavier than 230~$\text{M}_\sun$. Next, we derived a completeness-corrected age and mass function for our open cluster catalogue, including estimating that the Milky Way contains a total of $1.3 \times 10^5$ open clusters, only $\sim~4\%$ of which are currently known. Finally, we show that most open clusters have mass functions compatible with the Kroupa initial mass function.}
{We demonstrate Jacobi radii for distinguishing between bound and unbound star clusters, and publish an updated star cluster catalogue with masses and improved cluster classifications.}


   \maketitle

\section{Introduction}
\label{sec:introduction}




\paraphrased{Data releases from the \emph{Gaia} satellite have completely revolutionised the census of open clusters (OCs) \citep{cantat-gaudin_milky_2022}. Since the first full \emph{Gaia} data release \citep{brown_gaia_2018}, strides have been made in many aspects of the census, including ruling out many OCs reported before \emph{Gaia} as asterisms \citep{cantat-gaudin_clusters_mirages_2020,piatti_assessing_physical_2023,hunt_improving_2021,hunt_improving_open_2023}, detecting many thousands of new objects thanks to \emph{Gaia's} high-precision astrometry \citep[e.g.][]{liu_catalog_newly_2019,castro-ginard_hunting_open_2020}, and determining cluster parameters to higher levels of accuracy than previously possible \citep[e.g.][]{bossini_age_2019,cantat-gaudin_painting_2020}. Nevertheless, the OC census still has room for improvement, with a major issue being that current observational definitions of OCs do not seem to be robust enough to distinguish them from unbound moving groups (MGs) \citep{hunt_improving_open_2023}.}

\paraphrased{\cite{cantat-gaudin_clusters_mirages_2020} improve on the first major catalogue of OCs in the \emph{Gaia} era, \citep{cantat-gaudin_gaia_2018}, searching for additional OCs in \emph{Gaia} data and using a set of observational criteria to distinguish between plausible OCs and asterisms.} \rewrite{Their criteria are as follows: firstly, a candidate OC should be a clear overdensity, including that it has at least roughly ten member stars; secondly, it must have a colour-magnitude diagram (CMD) that follows a clear isochrone, indicating that a given cluster is a co-evolutionary population of stars with the same age and chemical composition; and finally, a candidate OC must pass criteria that distinguish asterisms that cannot physically be gravitationally bound from potential bound OCs: namely, a median radius $r_{50}$ less than 15~pc, and a proper motion dispersion that corresponds to an internal velocity dispersion smaller than 5~km\,s\textsuperscript{-1} (or 1~mas\,yr\textsuperscript{-1} for distant clusters where \emph{Gaia} measurement uncertainties are dominant.)}

\rewrite{The criteria on the density (or number of stars) and CMD quality of an OC candidate are common practice in the literature. For instance, \cite{froebrich_systematic_survey_2007}, \cite{cantat-gaudin_gaia_2019}, and \cite{hunt_improving_2021} (hereafter Paper~I) require a candidate new cluster to be a clear overdensity, while works such as \cite{platais_search_star_1998}, \cite{castro-ginard_new_2018}, and \cite{liu_catalog_newly_2019} are examples of works that use cluster CMDs to validate candidate new objects. The \cite{cantat-gaudin_clusters_mirages_2020} criteria on the possibility of a cluster being bound have also been adopted in the literature, with works such as \cite{hunt_improving_2021} and \cite{castro-ginard_hunting_open_2022} using them to validate new OC candidates.}

\paraphrased{In \cite{hunt_improving_open_2023} (hereafter Paper~II), we used \emph{Gaia} DR3 data \citep{gaiacollaboration_gaia_data_2022} to construct a large, homogeneous catalogue of star clusters}. However, despite our methodology being originally intended to only detect OCs, \paraphrased{many of the clusters we detected appear to be MGs, often having sparse or `flat' stellar distributions -- unlike the clustered appearance of canonically bound OCs such as the Pleiades.} Many of the suspected MGs we detected are consistent with being single populations of co-evolutionary stars based on our CMD classifier in Paper~II, and most of the MGs we detected still pass the observational criteria on the boundness of an OC proposed in \cite{cantat-gaudin_clusters_mirages_2020}. In Paper~II, we suggested that these criteria are too permissive to accurately classify the many sparse star clusters we are able to detect near to the Sun. The inability to distinguish precisely between bound and unbound clusters limits the scientific usability of catalogues such as the one in Paper~II, \paraphrased{with a significant proportion of the catalogue's content being likely to be MGs -- particularly within around 1~kpc from the Sun.}

\drafttwo{In this work, we aim to create a new way to distinguish between bound and unbound clusters, utilising a relationship between the mass and Jacobi radius of a self-gravitating cluster. With this method, we aim to classify all objects in Paper~II into bound and unbound objects, in addition to demonstrating the applicability of this method to future studies of the Milky Way's clusters, such as with upcoming data releases like \emph{Gaia} DR4.} Firstly, we provide an overview of some background theory in Sect~\ref{sec:theory}. In Sect.~\ref{sec:masses}, \paraphrased{we outline how we calculate cluster masses and radii, including how we correct for selection effects and unresolved binary stars.} Section~\ref{sec:results} outlines the results of this work, including how many clusters in Paper~II are bound OCs and how they are distributed. We explore the results of this work further in Sect.~\ref{sec:discussion}, including using our cluster masses to estimate the mass-dependent completeness of the \emph{Gaia} DR3 OC census, the age and mass functions of the OC census, and the compatibility of clusters with a Kroupa IMF \citep{kroupa_variation_initial_2001}. Section~\ref{sec:conclusion} concludes this work.

\section{Theoretical relations on the boundness of a cluster}
\label{sec:theory}

In this section, we review some theory on how the boundness of a star cluster could be measured. We aim to find relations that can be straightforwardly applied to \emph{Gaia} data of star clusters.

\subsection{The virial theorem}

One of the most common and widely used relations in astrophysics is the virial theorem, which states that \drafttwo{a system under the influence of gravitation and in equilibrium} should have twice as much kinetic energy $T$ as it has potential energy $U$, $2T = |U|$. For a star cluster with a distribution function following a \cite{plummer_problem_distribution_1911} model, this implies that a star cluster in virial equilibrium should have a one-dimensional velocity dispersion $\sigma_\text{1D}$ equal to an ideal virial velocity dispersion $\sigma_\text{vir}$ given by \citep{portegies_zwart_young_2010}:

\begin{equation}
	\sigma_\text{vir} = \sqrt{\frac{GM}{\eta r_\text{hm}}} \approx \sigma_\text{1D}\text{ for a bound cluster},
	\label{eqn:virial_velocity}
\end{equation}

\noindent where $G$ is the gravitational constant, $M$ is the cluster's mass, $r_{hm}$ is the cluster's half-mass radius, and $\eta$ is a constant equal to $\sim10$ for a typical cluster -- although it can be as low as $\sim5$ or as high as $\sim30$ depending on the cluster's spatial distribution \citep{portegies_zwart_young_2010}. This relation is used to analyse the dynamics of a small subset of nearby OCs in works including \cite{bravi_gaiaeso_survey_2018}, \cite{kuhn_kinematics_young_2019}, and \cite{pang_3d_morphology_2021}. 

\paraphrased{Equation~\ref{eqn:virial_velocity} may offer some explanation on why the individual empirical cuts presented in \cite{cantat-gaudin_clusters_mirages_2020} seem to be inadequate to distinguish between OCs and MGs from Paper~II. As an example, consider a small cluster with a radius of $r_{50}=1$~pc and a velocity dispersion of around 2~km\,s\textsuperscript{-1}. Equation~\ref{eqn:virial_velocity} predicts that this cluster would need a mass of $\sim10^4$~$\mathrm{M}_\sun$ to be virialised -- a value far higher than almost all OCs in the Milky Way, and clearly unrealistic for typical small MGs that we detect in Paper~II. Instead of adopting individual radius and velocity dispersion cuts, it appears that the expected radius and velocity dispersion of OCs must be `calibrated' individually based on a cluster's mass.}

However, during the preparation of this work, we found this relation impossible to apply successfully to all clusters in Paper~II. Velocity dispersions are easily contaminated by \emph{Gaia} measurement uncertainties, binary stars, unbound stars (including stars in cluster tidal tails), perspective expansion, and interloping field stars. \drafttwo{For instance, we found that binary stars (resolved or unresolved) often contribute 500~m\,s\textsuperscript{-1} or more to cluster velocity dispersions derived using proper motions, and incorrect removal of tidal tails can contribute as much as 2~km\,s\textsuperscript{-1} in the worst cases. In addition, \emph{Gaia} measurement uncertainties become dominant in the proper motion dispersion of most clusters above a few kpc, making it difficult to make a meaningful measurement of $\sigma_\text{1D}$ for many clusters.} Accounting for all of these effects for all clusters in Paper~II and arriving at accurate measurements of $\sigma_\text{1D}$ was not possible. In addition, it is predicted theoretically that star clusters are often supervirial, such as during phases of expansion for young clusters \citep{banerjee_how_can_2017, krause_physics_2020} -- making it difficult to make a scientifically motivated cut on $\sigma_\text{vir}$ in many cases, as the velocity dispersion of many bound clusters can be expected to be supervirial, with measurements in multiple studies supporting this hypothesis \citep[e.g.][]{bravi_gaiaeso_survey_2018,kuhn_kinematics_young_2019,pang_3d_morphology_2021}.

\subsection{Jacobi radii}

Using currently available data, we found it was much more successful to only rely on cluster masses and radii alone to distinguish between OCs and MGs. An instantaneously bound cluster should have a Roche surface, within which its potential is stronger than that of its host galaxy. In principle, a cluster with no radius at which its potential is stronger than that of the Milky Way will have no Roche surface, and is hence not self-gravitating. The Roche surface of a given cluster can be measured by considering its Jacobi radius, $r_J$, which is the distance from the centre of a cluster to its $L_1$ Lagrange point. This is given by \citep{portegies_zwart_young_2010, ernst_simulations_hyades_2011}:

\begin{equation}
	r_J = \left( \frac{GM}{4\Omega^2 - k^2} \right)^{\frac{1}{3}},
	\label{eqn:jacobi_radius}
\end{equation}

\noindent
\paraphrased{which relates $r_J$ to the mass $M$ of a cluster, weighted by the circular frequency $\Omega$ and epicyclic frequency $k$ of the cluster's orbit around its host galaxy, assuming that the orbit is circular. Outside of $r_J$, the host galaxy's potential is dominant -- such as for stars in the tidal tails of a cluster, which are no longer bound to their parent cluster \citep{meingast_extended_2021}. Assuming that a cluster fills its Roche surface, $r_J \approx r_t$ from a \cite{king_structure_star_1962} model fit \citep{binney_galactic_dynamics_1987}, with this relationship being used by works such as \cite{piskunov_tidal_radii_2008} to derive the masses of OCs based on their size.}

Despite the fact that some open clusters, such as the Hyades, have been shown to have higher than expected stellar velocity dispersions and are supervirial and dissolving, these dissolving clusters are nevertheless dense enough to be currently self-gravitating \citep{oh_kinematic_modelling_2020,meingast_extended_2021}. The same applies for young clusters that have recently been observed to be in possible supervirial expansion phases not long after their initial formation \citep{kuhn_kinematics_young_2019}. On the other hand, MGs of all kinds (including sparse OB associations) are not bound, and are actively dissolving into the disk, meaning that they should have no radius at which they have a Roche sphere, which should be possible to measure with Eqn.~\ref{eqn:jacobi_radius}. For the remainder of this work, we aim to apply this equation to distinguish between bound and unbound star clusters.

\section{Mass and radius calculations}
\label{sec:masses}

\paraphrased{In this section, we describe how we calculated photometric masses and Jacobi radii for all clusters within 15~kpc from Paper~II. Much of this method was originally described in \cite{hunt_improving_census_2023}, but is outlined again here to aid in the reading of this work. This method contains five steps that we discuss in the following subsections. Firstly, we derived the photometric masses of member stars in every cluster. Next, we corrected for selection effects. We then applied a correction for unresolved binary stars. Mass functions were then fitted and integrated to calculate total cluster mass. Finally, this process was repeated at different radii to find the Jacobi radius of each cluster.}


\subsection{Calculation of stellar masses}
\label{sec:masses:isochrones}

\begin{figure}[t]
    \centering
    \includegraphics[width=\columnwidth]{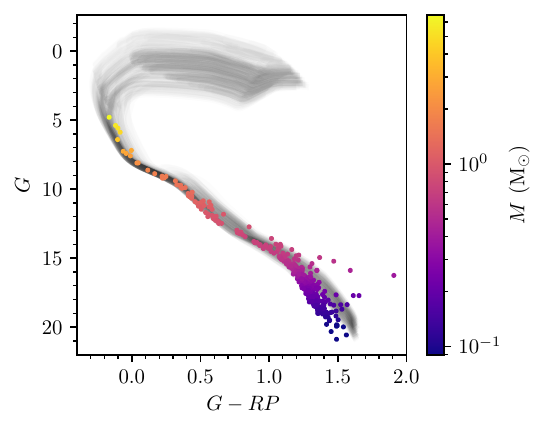}
    \caption{\paraphrased{CMD of member stars of NGC~2451A shaded by their calculated stellar mass. 100 sampled isochrones for the cluster from Paper~II are shown in black. \citep[Adapted from][]{hunt_improving_census_2023}}}
    \label{fig:masses:stellar_masses}
 \end{figure}

\paraphrased{Following a similar method to that used by works including \cite{meingast_extended_2021} and \cite{cordoni_photometric_binaries_2023}, we began by using PARSEC \citep{bressan_parsec_stellar_2012} isochrone fits from Paper~II to estimate the masses of member stars of each cluster.} \paraphrased{To calculate stellar masses, we used the predicted mass of stars as a function of $G$-band magnitude from our fitted Paper~II isochrones, $m(G)$, which was accurate for most cluster members. However, the oldest clusters in our sample often contain evolved giant stars whose $G$-band magnitude is less than the tip of the main sequence in the cluster -- meaning that $m(G)$ is not a one-to-one mapping from magnitude to mass for some cluster members. Therefore, in regions where our fitted PARSEC isochrones do not have a one-to-one mapping from magnitude to mass,} \referee{we also used $BP-RP$ colour indices to decide on the best stellar mass for a given cluster member. We elected not to use the $BP-RP$ index of most stars as $BP$ and $RP$ magnitudes are frequently underestimated for very red or blue stars with magnitudes $G\gtrsim19$ \citep{riello_gaia_early_2021}. The regions where we do use $BP-RP$ colour indices were not within ranges where $BP$ and $RP$ are underestimated, however, due to the blue colour of these regions and due to our studied clusters being within 15~kpc.}

\paraphrased{To incorporate the uncertainty on our isochrone fits from Paper~II, we repeated this process 100 times for 100 sampled isochrones from our variational inference neural network in Paper~II. This incorporated uncertainties on the age, extinction, and distance to stars into our mass estimates for them. Figure~\ref{fig:masses:stellar_masses} illustrates this process for NGC~2451A, showing 100 sampled isochrones from Paper~II and our estimated stellar masses for each star with the shading of points.}

\referee{It is worth discussing further the limitations and assumptions of this method. Firstly, since our Paper~II photometric parameters do not include metallicities, our masses are biased for clusters with particularly low or high metallicities. To quantify this systematic, we repeated our entire pipeline on 143 randomly selected clusters but assuming metallicities of $[\text{Fe}/\text{H}]=+0.5$ and $-0.5$~dex, testing how masses change given $[\text{Fe}/\text{H}]$ values at the upper and lower limit of those observed in OCs \citep{kharchenko_global_2013,bossini_age_2019}. Assuming a high metallicity increases masses by an average of $+7\%$, while a low metallicity decreases masses by an average of $-12\%$. The mean metallicity of OCs is approximately solar \citep{kharchenko_global_2013}, and so these values are edge-case limits that will mostly impact clusters at particularly high or low galactocentric radii which are most likely to have non-solar metallicities \citep{spina_mapping_2022}. In the future, it will be important to include spectroscopic metallicity estimates in machine learning OC parameter inference to improve the accuracy of OC masses further.}

\paraphrased{Next of note is that the effects of binary stars were not included in our interpolation scheme. Our stellar mass estimates are hence only estimates of the mass of the primary star in any binary system. To mitigate this effect, we applied a correction to the overall cluster mass function for unresolved binaries in Sect.~\ref{sec:masses:binaries}.}

\referee{Finally, our use of PARSEC isochrones also influences our derived stellar masses, and our mass estimates may differ to those derived using other stellar evolution models. We investigated how using MIST isochrones \citep{choi_mesa_isochrones_2016} would change our mass estimates, using limited comparisons between our fitted PARSEC isochrones and MIST isochrones at the same age. $m(G)$ for PARSEC and MIST isochrones is generally very similar for $m\gtrsim0.8$~M$_\sun$ at all ages, and hence $m(G)$ for clusters at distances greater than 1~kpc (where almost all observed stars are greater than this mass) will be similar. We estimate that MIST-derived total cluster masses would still be lower than PARSEC ones for such clusters, although no more than $\sim5\%$ lower. However, $m(G)$ is appreciably lower in MIST compared to PARSEC for stars with masses below $\sim0.8$~M$_\sun$ at all ages, meaning that low mass stars would be assigned lower masses by MIST isochrones given the same brightness. We estimate that this would result in total cluster masses that are at most $\sim15\%$ lower for clusters within 300~pc (which have mass functions most strongly impacted by the low-mass stars where MIST and PARSEC isochrones have the weakest agreement.)}


\subsection{Correction for selection effects}
\label{sec:masses:selection}

\begin{figure}[t]
    \centering
    \includegraphics[width=\columnwidth]{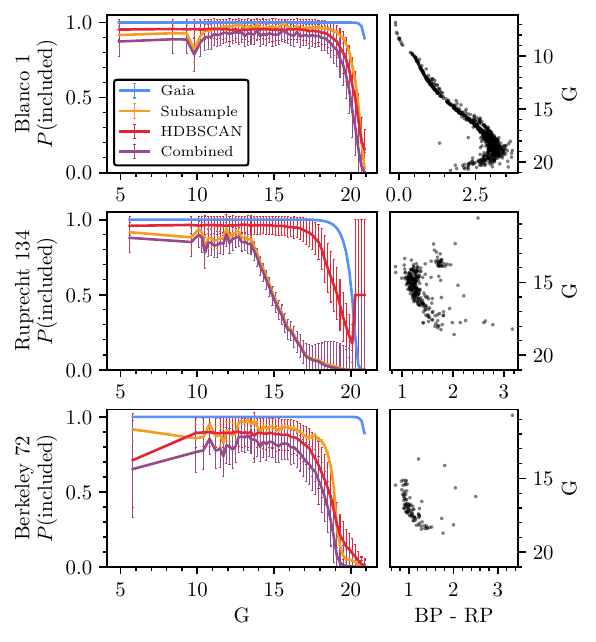}
    \caption{\paraphrased{Computed cluster selection functions for Blanco~1 \emph{(top row)}, Ruprecht~134 \emph{(middle row)}, and Berkeley~72 \emph{(bottom row)}. The left panel in each row shows our adopted \emph{Gaia} (blue), subsample (orange), and algorithm (HDBSCAN, red) selection functions as a function of magnitude for each cluster, in addition to the multiplicative total selection function (purple). The CMD of each cluster is shown for reference on the right panels. \citep[Adapted from][]{hunt_improving_census_2023}}}
    \label{fig:masses:selection_effects}
\end{figure}

\referee{Although our Paper~II cluster membership lists aimed to be as complete as possible, with membership lists including stars as faint as $G\sim20$, there remain a number of selection effects that limit the completeness of our membership lists that must be accounted for to derive accurate cluster masses. From inspection of the CMDs of clusters that are challenging to recover, such as those in regions of high crowding where \emph{Gaia} data becomes incomplete \citep{gaiacollaboration_gaia_early_2021} or distant clusters where our adopted clustering technique can miss member stars (Paper~II), there are clearly selection effects that would otherwise influence our derived mass functions. In this subsection, we describe how we model the selection effects that impact each of our cluster membership lists. We refer readers to \cite{hunt_improving_census_2023} for more detail on our method.} 

\referee{We consider three different effects that could result in a real star to be missing from our membership lists. Firstly, there is the probability that a given star with parameters $\vec{q}$ appears in the \emph{Gaia} DR3 catalogue of 1.8~billion sources, $S_{\text{Gaia}}(\vec{q})$. Then, there is the conditional probability that a source in \emph{Gaia} DR3 was included in the subset of \emph{Gaia} data that we used for clustering analysis, which is all 729~million stars with a full astrometric solution, $BP$ and $RP$ photometry, and a \cite{rybizki_classifier_spurious_2022} v1 quality flag of greater than 0.5, $S_{\text{subsample}}(\vec{q} \mid \vec{q}\text{ in Gaia})$. Finally, there is the additional probability that our adopted clustering algorithm in Paper~II, HDBSCAN \citep{hutchison_hdbscan_2013,mcinnes_hdbscan_hierarchical_2017}, assigns this star as a member of the cluster, $S_{\text{algorithm}}(\vec{q} \mid \vec{q}\text{ in subsample})$ -- which is decreasingly likely depending on how clearly separated a cluster is from the surrounding field. These effects are multiplicative \citep{rix_selection_functions_2021, castro-ginard_estimating_selection_2023}, hence giving an overall probability $S_{\text{cluster}}(\vec{q})$ that a star with parameters $\vec{q}$ appears in our adopted cluster membership list:}

\begin{equation}
    \begin{split}
        S_{\text{cluster}}(\vec{q}) = S_{\text{Gaia}}(\vec{q}) \cdot S_{\text{subsample}}(\vec{q} \mid \vec{q}\text{ in Gaia}) \\ \cdot S_{\text{algorithm}}(\vec{q} \mid \vec{q}\text{ in subsample}).
    \end{split}
\end{equation}

\referee{The first two terms, $S_{\text{Gaia}}(\vec{q})$ and $S_{\text{subsample}}(\vec{q} \mid \vec{q}\text{ in Gaia})$, are calculated directly from the works of \cite{cantat-gaudin_empirical_model_2023} and \cite{castro-ginard_estimating_selection_2023}. In the first work, \cite{cantat-gaudin_empirical_model_2023} derive an empirical probability that a source appears in \emph{Gaia} DR3 by comparing the \emph{Gaia} dataset to photometric surveys deeper than \emph{Gaia} itself. They describe the probability that a source is included in \emph{Gaia} based on its position, which is a good tracer for the extent of crowding in a given region, in addition to its $G$-band magnitude, which is a strong predictor of how well it would be processed by the \emph{Gaia} telescope and data processing pipeline. Values of $S_{\text{Gaia}}(\vec{q})$ as a function of position and magnitude were queried directly from the \texttt{gaiaunlimited} Python package\footnote{\url{https://github.com/gaia-unlimited/gaiaunlimited}} \citep{cantat-gaudin_empirical_model_2023}.}

\referee{Next, \cite{castro-ginard_estimating_selection_2023} outline a method to determine the probability that a source appears in a given subsample of the \emph{Gaia} dataset $S_{\text{subsample}}(\vec{q} \mid \vec{q}\text{ in Gaia})$, using a method from \cite{rix_selection_functions_2021}. We implemented the empirical \cite{castro-ginard_estimating_selection_2023} method as a function of position and $G$-band magnitude alone, which we found to be good predictors of the probability of a source being in our adopted subsample of the \emph{Gaia} dataset. The subsample of \emph{Gaia} data we used in Paper~II was largely to restrict our analysis to only sources with a good-quality astrometric solution, which is strongly influenced by the position and brightness (S/N) of a source. Since the method in \cite{castro-ginard_estimating_selection_2023} bins sources to calculate $S_{\text{subsample}}(\vec{q} \mid \vec{q}\text{ in Gaia})$, we selected all stars in the on-sky region a given cluster covers and binned them by $G$-band magnitude in bins of size 0.2~mag. To prevent under-sampling of bins for bright sources, bins were merged until every bin contained at least ten stars.}

\referee{Finally, to model the impact of incompleteness due to the clustering algorithm we used in Paper~II, we developed a stochastic technique to simulate the probability of a true member star of a given cluster being assigned as a member by the algorithm. The chance of a star being correctly assigned as a cluster member depends strongly on its astrometric precision (Paper~II). For a star with less astrometric precision, its position in the five dimensional \emph{Gaia} astrometry that we performed clustering for will be further away from the centre of a cluster, meaning that it is more likely to be missed by our clustering algorithm. This is particularly the case for distant clusters, where \emph{Gaia} uncertainties are often larger than the true underlying parallax or proper motion dispersion of a cluster. Since the clusters in this work are generally no smaller than 0.1$^\circ$ in angular extent, astrometric errors on star position in \emph{Gaia} DR3 are negligible compared to the size of clusters, and so this effect is only dependent on the proper motion and parallax precision of cluster members.}

\referee{This effect was modelled by performing simulations of whether or not stars with simulated astrometry would appear within a cluster. For every cluster, we simulated 100\,000 stars with magnitudes uniformly distributed in the range $2<G<21$. Astrometric errors were assigned to each star by randomly selecting stars with a similar magnitude in the vicinity of the real cluster and using their errors directly. Then, ten random samplings of the proper motions and parallaxes of each star were performed. To estimate whether or not each simulated star would be assigned as a member of a given cluster, we fitted a 3D ellipsoid to the proper motions and parallaxes of our Paper~II cluster members for each cluster, and then calculated how often each simulated star appeared within each fitted ellipsoid to calculate the probability that a given star was included in our membership list.}

\referee{The estimated selection functions and the CMDs of three OCs are shown in Fig.~\ref{fig:masses:selection_effects}, showing how different OCs have CMDs dominated by different effects. In the first case, Blanco~1 is a high galactic latitude, nearby ($d=240$~pc) cluster that is easy to detect and is clearly separated from the field. It has a CMD that is visually well-populated to magnitudes fainter than even $G\sim20$. This is reflected in its estimated selection function, which is largely complete for $G<19$. On the other hand, Ruprecht~134 is a cluster in an extremely crowded field near the galactic centre ($d=2.3$~kpc), being one of the most incomplete clusters in our Paper~II catalogue. It is strongly impacted by the selection effect of our subsample, which removes a large number of sources with anomalous astrometry due to crowding -- in addition to the selection function of \emph{Gaia} DR3, which reduces sharply at $G \sim 20$ in this region. Finally, Berkeley~72 is a more distant cluster ($d=5.1$~kpc). Owing to its distance and relative sparsity, its selection function is mostly dominated by the selection function of our clustering algorithm, although the subsample selection function also makes a contribution due to the cluster's location in a somewhat crowded region of the galactic disk. From these three examples alone, it is clear that all three selection effects impact every cluster in different ways that must all be considered.}

\referee{In addition, it is worth noting that no cluster is estimated to be 100\% complete at any magnitude. We suggest that many of these potentially missing stars are likely to be multiple stars. During the processing of \emph{Gaia} DR3, all stars were assumed to be single; however, binaries with large deviations from ideal single-star astrometry will have higher errors in their astrometric fits \citep{lindegren_gaia_early_2021a}, and are less likely to appear in the subsample of stars with good-quality astrometry that our Paper~II catalogue was constructed from.}

\subsection{Correction for unresolved binaries}
\label{sec:masses:binaries}

\paraphrased{The next step in our method corrected for unresolved binary stars. Since our inferred stellar masses in Sect.~\ref{sec:masses:selection} assumed that stars are single, an additional correction for unresolved binaries is important to avoid our final cluster masses being biased to low values.}

\paraphrased{Ideally, it would be possible to directly detect all binaries in a given cluster and measure the mass ratio $q$ of each binary system, using this as a correction to each star's estimated mass. However, such direct measurements are not possible based on \emph{Gaia} DR3 data alone, and particularly not for all 7167 clusters in Paper~II. Some works have recently studied the binary star fraction in a subset of reliable OCs, including \cite{cordoni_photometric_binaries_2023} who measure it for 78 OCs and \cite{donada_multiplicity_fraction_2023a} who measure it for 202 OCs within 1.5~kpc. Nevertheless, both works are only able to measure the binary fraction for mass ratios $q \gtrsim 0.6$, due to the difficulty of distinguishing between low-mass ratio binary stars and single stars on the main sequence, especially in the presence of differential reddening. Particularly since the binary star fraction in the solar neighbourhood has been tentatively shown to peak at $q \sim 0.3$ for most stars below 2 to 5 M$_{\sun}$ \citep{moe_mind_your_2017}, meaning that many stars are most likely to have binaries with $q$ below values that can be measured with \emph{Gaia} DR3, we conclude that robust direct measurements of the mass ratio of most binaries in OCs are not possible, and instead used an approximate correction to our adopted stellar masses that accounts for binaries at all $q$ values.}

\paraphrased{We derived corrections to apply to our final cluster mass functions by using the selection effect corrected multiplicity fraction, companion star frequency, and mass ratio distribution of field stars from \cite{moe_mind_your_2017}. Binary stars for each cluster were simulated based on these distributions. To simulate whether a binary is resolved, which is frequently the case for nearby clusters \citep{donada_multiplicity_fraction_2023a}, the period and eccentricity distributions from \cite{moe_mind_your_2017} were used to simulate the mean separation of each simulated binary, which was then compared to the angular resolution of \emph{Gaia} DR3 \citep{gaiacollaboration_gaia_early_2021}. Depending on the distance to a cluster and the mass range of the mass function bin to be corrected, this binary star correction increases mass bins by 10\% to 30\%, while also inflating our quoted uncertainties on cluster masses significantly, owing to the approximate nature of this method. We estimate that in the worst cases, errors due to our assumed field-like binary star population could contribute additional systematics of up to $\sim20\%$ on our final cluster masses. In the future, it will be important to improve methods to determine which stars are binaries to improve the accuracy of OC masses further.}

\subsection{Mass function fits}
\label{sec:masses:imf_fits}

\begin{figure}[t]
    \centering
    \includegraphics[width=\columnwidth]{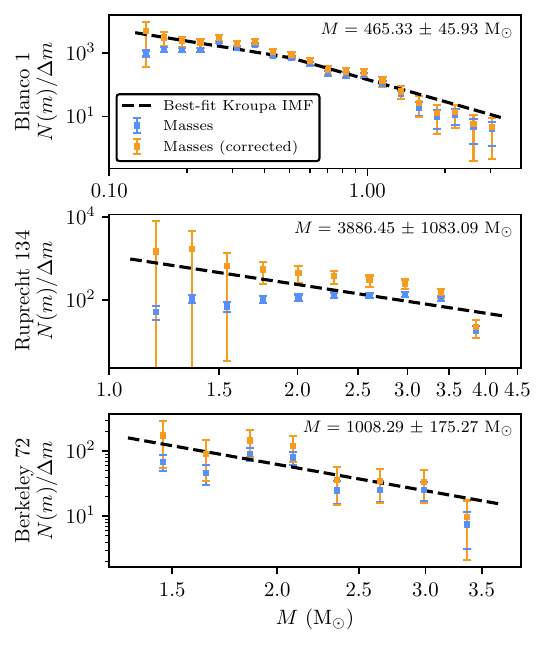}
    \caption{\paraphrased{Mass functions for the three OCs from Fig.~\ref{fig:masses:selection_effects}. Original binned stellar masses are shown by the blue squares, while binned masses corrected for selection effects and unresolved binary stars are shown by the orange squares. The dashed black line shows our fitted Kroupa IMF, with our calculated total cluster mass and corresponding uncertainty in the top right. \citep[Adapted from][]{hunt_improving_census_2023}}
    }
    \label{fig:masses:mass_functions}
\end{figure}

\paraphrased{The final step in our cluster mass measurement pipeline was to fit a mass function to each cluster and integrate it to derive a total cluster mass, including stars too faint to be observed.} \paraphrased{There are a number of different functional forms for mass functions that can be adopted \citep{krause_physics_2020}, with a common form being a broken power law with a break point at a value like 0.5~M$_\sun$ \citep{kroupa_variation_initial_2001}. Some works, such as \cite{cordoni_photometric_binaries_2023}, derive OC masses while fitting bespoke broken power law mass functions to every cluster. However, since most clusters in our sample are more distant than 1~kpc, or have $A_V \gtrsim 2$, they contain few or no stars below the typical mass function break point of $\sim0.5$~M$_\sun$, making it impossible to fit two-part mass functions to them. Extrapolating single power law mass functions measured for high mass stars down to lower stellar masses would cause severe over-estimates of our cluster masses in these cases, since it has been robustly measured that clusters form with mass functions that are significantly less steep at masses of around 0.5~M$_\sun$ and below \citep{krause_physics_2020}. Instead, we adopted a `safer' approach, and fit only a \cite{kroupa_variation_initial_2001} IMF (hereafter Kroupa IMF) to every cluster.}

\paraphrased{To fit cluster mass functions, we used the \texttt{imf} Python package\footnote{\url{https://github.com/keflavich/imf}} and performed least squares fitting of the amplitude of each cluster mass function after correcting for selection effects and unresolved binary stars. In general, we found that the majority of clusters had mass functions well approximated by a Kroupa IMF, albeit only after correcting masses for selection effects and binary stars. Figure~\ref{fig:masses:mass_functions} shows mass functions for the three clusters from Fig.~\ref{fig:masses:selection_effects}, all of which have slopes well approximated by Kroupa IMFs after incorporating corrections. Section~\ref{sec:discussion:kroupa_imf} compares our cluster mass functions to the Kroupa IMF further.}

\referee{Finally, to convert our fitted IMF into a total cluster mass, each fitted IMF was integrated from a lower limit of 0.03~M$_\sun$ to the highest observed stellar mass in the cluster. This lower limit is slightly lower than the 0.08~M$_\sun$ lower limit used in some other works \citep[e.g.][]{meingast_extended_2021} which corresponds to the minimum mass at which nuclear fusion still occurs. Our lower limit of 0.03~M$_\sun$ intentionally also includes brown dwarfs, which are also observed in OCs \citep{moraux_brown_dwarfs_2003} -- but stops short of integrating from 0~M$_\sun$, as companion objects around stars with masses below 0.03~M$_\sun$ are often considered planets and have a poorly constrained IMF \citep{akeson_nasa_exoplanet_2013}, and the quantity of these objects that are free-floating is also poorly constrained. Nevertheless, the choice of lower limit makes a negligible difference on the final cluster mass on the order of $\sim1\%$.}

\subsection{Jacobi radius inference}
\label{sec:masses:jacobi}

\paraphrased{The last step of our method was to calculate the mass of each cluster at all radii, which was then compared against the theoretically predicted Jacobi radius of a cluster of that mass and radius. This produced a probability that a given cluster has some radius $r_J$ at which its gravitational potential is stronger than that of the Milky Way, hence measuring whether a given cluster is self-gravitating and (currently) bound.}

\paraphrased{Firstly, we repeated our mass measurement pipeline at all cluster radii, deriving cluster mass as a function of cluster radius $M_\text{obs}(r)$. We did not consider cluster radii where a cluster had fewer than ten member stars, as OCs are usually defined to contain at least ten member stars to differentiate them from multiple star systems \citep{cantat-gaudin_clusters_mirages_2020, portegies_zwart_young_2010}. In addition, we calculated the total cluster mass including all assigned member stars (such as tidal tails) $M_A$.}

\paraphrased{Next, we used the method of \cite{meingast_extended_2021} to calculate the theoretical Jacobi mass as a function of radius for each cluster, $M_J(r)$, by inverting Eqn.~\ref{eqn:jacobi_radius}. One must assume a model of the galactic potential to calculate $\Omega$ and $k$ within Eqn.~\ref{eqn:jacobi_radius}, for which we used the \texttt{galpy} \texttt{MWPotential2014} model of the Milky Way's potential \citep{bovy_galpy_python_2015}.} \referee{This potential is smooth, and does not include spiral arms or giant molecular clouds (GMCs), but was fit to a wide variety of data and should be accurate enough for our circumstances. In practice, $r_J$ depends relatively weakly on the assumed potential model, due to its cube root dependence on the quantity $\sqrt[3]{4 \Omega^2 - k^2}$ calculated from the potential. Within our adopted potential model, this quantity is only a factor of four larger between clusters at the lowest galactocentric radii in this study ($\sim$2~kpc) and the highest ($\sim$20~kpc). In addition, we are most interested in this work in distinguishing between MGs and OCs in the solar neighbourhood, for which we expect the galactic potential to be most well determined by this model. These frequencies could of course be less accurate at higher distances from the Sun, for which the galactic potential is not as well constrained.} 

\referee{Nevertheless, the smoothness of our adopted potential model could be a source of bias. Due to their increased gas density relative to the rest of the galaxy, GMCs and spiral arms will have a stronger potential, with collisions with GMCs and spiral arms being major contributors to mass loss and destruction of OCs \citep{krause_physics_2020}. Given the weak dependence of Eqn.~\ref{eqn:jacobi_radius} on our assumed potential model, this effect is likely to be small for spiral arms, which have been measured to have around a 50\% increase in gas mass \citep{colombo_sedigism_survey_2022} -- which results in only a small change in local potential, as the local potential acting on an OC in a spiral arm will still be dominated by the Milky Way's (assumed) smooth dark matter halo \citep{bovy_galpy_python_2015,cautun_milky_way_2020}. However, due to their significantly higher gas density, the potential in a typical GMC will be significantly higher \citep{krause_physics_2020}. Since we are most interested in classifying suspicious candidate new OCs in the solar neighbourhood, it is somewhat fortunate that the Sun is within a bubble that contains little gas \citep{zucker_solar_neighborhood_2023}. As an additional check, we matched our catalogue against the catalogue of nearby molecular clouds in \citep{cahlon_parsecscale_catalog_2024}. Only one cluster (HSC~598) is within 25~pc of a molecular cloud, at a separation of 8~pc. However, this cluster is already classified as an MG by our pipeline below, and so the influence of its neighbouring molecular cloud on our final results is negligible. Future work using deeper datasets (such as \emph{Gaia} DR4 or DR5) will presumably work with a deeper catalogue that includes more clusters out to larger distances, and hence the influence of GMCs and spiral arms on the local potential surrounding clusters may have to be considered.}

\paraphrased{Finally, with theoretical values of $M_J(r)$ calculated for each cluster, $M_J(r)$ and $M_\text{obs}(r)$ were compared to identify a possible Jacobi radius for each cluster. If a cluster has some radius $r$ at which the enclosed mass within this radius $M_\text{obs}(r)=M_J(r)$, then $r$ is taken as the cluster's Jacobi radius $r_J$, with corresponding enclosed mass $M_J$. In cases where $M_\text{obs}(r)<M_J(r)$ at all radii, the cluster is considered to have no valid Jacobi radius and is an MG. In addition, some clusters have $M_\text{obs}(r)>M_J(r)$ at all radii, meaning that the observed cluster is smaller than the Jacobi radius of the cluster given its observed mass. This is the case for some distant or difficult to detect clusters, for which it is likely that we only observe the innermost parts of the cluster, and do not detect stars out to the true cluster $r_J$. In these cases, we used the theoretical $r_J$ of all observed stars in the cluster as the cluster's $r_J$, although such values probably underestimate the true cluster $r_J$.}

\begin{figure}[t]
    \centering
    \includegraphics[width=\columnwidth]{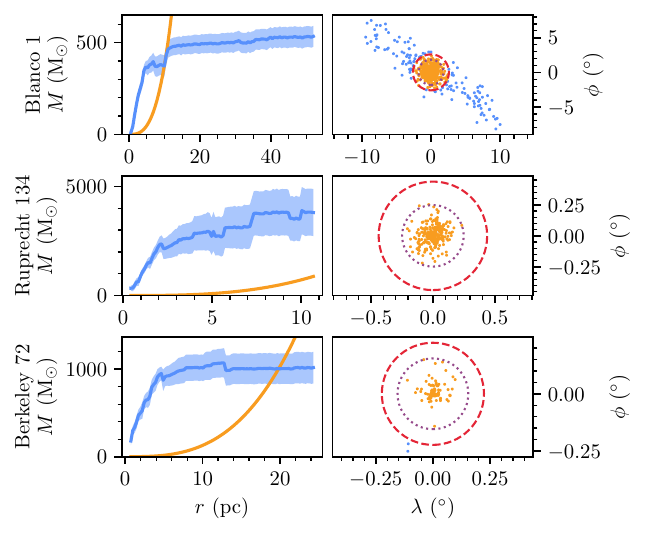}
    \caption{\referee{Jacobi radius calculation method shown for the three OCs from Fig.~\ref{fig:masses:selection_effects}, with each row corresponding to each OC. Within each row, the left panel shows the cluster mass as a function of radius from the centre of the cluster, where the blue line is our calculated total cluster mass with a shaded uncertainty region, and the orange line is the theoretical Jacobi mass for a cluster of that size given Eqn.~\ref{eqn:jacobi_radius}. The intersection of these lines is the cluster's $r_J$. The right panel in each row shows the cluster in an arbitrary coordinate frame centred on the cluster centre. Member stars within $r_J$ are shown in orange, with member stars outside of $r_J$ shown in blue. $r_J$ is indicated by the dashed red line. The dotted purple line denotes our approximate calculated \cite{king_structure_star_1962} tidal radius for each cluster from Paper~II.} \paraphrased{\citep[Adapted from][]{hunt_improving_census_2023}}
    }
    \label{fig:masses:radii_examples}
\end{figure}

\paraphrased{$M_\text{obs}(r)$ and $M_J(r)$ are shown in Fig.~\ref{fig:masses:radii_examples} for Blanco~1, Ruprecht~134, and Berkeley~72. As would be expected for these reliable clusters, all of them clearly have radii at which their enclosed mass is higher than the theoretical Jacobi mass, implying that they are self-gravitating bound clusters. As previously discussed, Ruprecht~134 is a difficult to detect cluster, for which we find $M_\text{obs}(r)>M_J(r)$ at all radii, implying that additional member stars in the outskirts of this cluster are yet to be detected.}

\begin{figure}[t]
    \centering
    \includegraphics[width=\columnwidth]{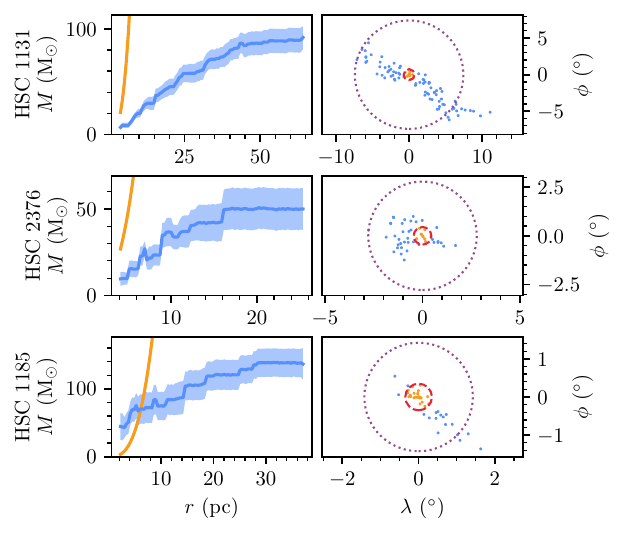}
    \caption{\paraphrased{Same as Fig.~\ref{fig:masses:radii_examples}, but for three candidate new OCs from Paper~II: HSC~1131 \emph{(top row)}, HSC~2376 \emph{(middle row)}, and HSC~1185 \emph{(bottom row)}. Although HSC~1131 and HSC~2376 do not appear to have a Jacobi radius, their most likely Jacobi radius is still shown in the plots in the right column. \citep[Adapted from][]{hunt_improving_census_2023}}
    }
    \label{fig:masses:radii_examples_sus}
\end{figure}

\paraphrased{However, since we are interested in using Jacobi radii to distinguish between bound and unbound clusters, the performance of this method on suspect clusters is most relevant. In the penultimate section of Paper~II, we highlighted three example candidate new OCs -- two of which appeared particularly suspicious due to their on-sky distributions not showing a clear `clumpy' cluster core as would be expected for an OC \citep{king_structure_star_1966}. Figure~\ref{fig:masses:radii_examples_sus} shows the mass as a function of radius of these three clusters. The two clusters that we highlighted as looking unlike OCs, HSC~1131 and HSC~2376, have $M_\text{obs}(r)<M_J(r)$ at all radii, strongly suggesting that they are indeed unbound MGs. On the other hand, HSC~1186, a cluster that does have a small central clump, appears compatible with being a small bound OC with a bound mass of $M\sim65$~M$_\sun$, as was also expected in Paper~II.}

\paraphrased{In total, it took less than 10\% of the total CPU wall time to perform mass and radius calculations as our original clustering analysis in Paper~II took. Not only does this method appear viable for distinguishing between OCs and MGs, it is also not especially computationally challenging, meaning that it could feasibly be incorporated into any future cluster blind searches.}

\section{Results}
\label{sec:results}

\begin{figure}[t]
    \centering
    \includegraphics[width=\columnwidth]{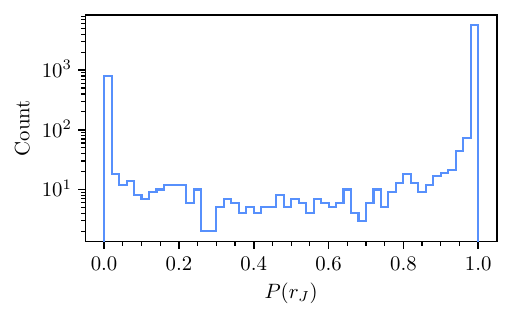}
    \caption{\paraphrased{Distribution of $P(r_J)$ for all clusters in this work, which is the probability that a cluster has a valid Jacobi radius. \citep[Adapted from][]{hunt_improving_census_2023}}}
    \label{fig:results:jacobi_radii_distribution}
\end{figure}

\begin{table*}[t]
    \caption{\label{tab:catalogue_clusters}Catalogue of star clusters with masses, object classifications, and Jacobi radii.}
    \centering
    \begin{tabular}{lccccccccc}
        \hline\hline
        Name & Type & $N_J$ & $N$ & $P(r_J)$ & $r_{50,J}$ (pc) & $r_J$ (pc) & $M_J$ (M\textsubscript{$\odot$}) & $M_A$ (M\textsubscript{$\odot$}) \\
        \hline
        \multicolumn{9}{c}{$\cdot \cdot \cdot$} \\ 
        
        Blanco 1 & \texttt{o} & 710 & 841 & 1.00 & 2.97 & 10.65 & 465.33 (45.93) & 529.09 (57.67) \\
        HSC 1142 & \texttt{m} & - & 159 & 0.00 & - & - & - & 109.73 (20.02) \\
        HSC 180 & \texttt{m} & - & 24 & 0.00 & - & - & - & 24.83 (5.35) \\
        HSC 2068 & \texttt{m} & - & 28 & 0.01 & - & - & - & 14.87 (4.32) \\
        HSC 2327 & \texttt{m} & 13 & 91 & 0.99 & 1.83 & 3.04 & 10.75 (3.79) & 64.43 (13.20) \\
        HSC 242 & \texttt{m} & - & 62 & 0.00 & - & - & - & 36.65 (6.45) \\
        HSC 2603 & \texttt{m} & - & 26 & 0.00 & - & - & - & 17.67 (4.35) \\
        HSC 2907 & \texttt{o} & 229 & 349 & 1.00 & 3.07 & 6.96 & 132.83 (13.41) & 184.66 (22.60) \\
        HSC 719 & \texttt{m} & - & 29 & 0.00 & - & - & - & 16.69 (2.32) \\
        HSC 782 & \texttt{m} & 22 & 47 & 1.00 & 0.84 & 3.30 & 13.66 (3.28) & 25.90 (6.99) \\
        IC 2391 & \texttt{o} & 316 & 376 & 1.00 & 1.98 & 7.66 & 169.43 (25.39) & 203.89 (27.55) \\
        IC 2602 & \texttt{o} & 440 & 638 & 1.00 & 3.46 & 8.52 & 237.25 (32.45) & 344.16 (42.26) \\
        Mamajek 2 & \texttt{o} & 98 & 226 & 1.00 & 3.10 & 6.10 & 90.56 (4.21) & 205.71 (14.24) \\
        Melotte 20 & \texttt{o} & 738 & 938 & 1.00 & 4.28 & 10.30 & 391.62 (54.11) & 502.80 (65.91) \\
        Melotte 22 & \texttt{o} & 1639 & 1721 & 1.00 & 3.61 & 13.69 & 946.51 (86.77) & 984.55 (92.36) \\
        Melotte 25 & \texttt{o} & 569 & 927 & 1.00 & 4.08 & 8.06 & 193.04 (42.64) & 409.00 (56.95) \\
        NGC 2632 & \texttt{o} & 1224 & 1314 & 1.00 & 3.95 & 13.73 & 945.01 (72.73) & 1012.01 (73.71) \\
        OCSN 49 & \texttt{m} & - & 265 & 0.31 & - & - & - & 218.82 (23.91) \\
        Platais 10 & \texttt{o} & 58 & 197 & 1.00 & 2.50 & 4.74 & 42.94 (8.73) & 166.79 (19.77) \\
        UPK 612 & \texttt{m} & 28 & 228 & 1.00 & 1.57 & 3.20 & 28.60 (6.28) & 112.79 (28.61) \\
        
        \multicolumn{9}{c}{$\cdot \cdot \cdot$} \\ 
        \hline
    \end{tabular}
    \tablefoot{Shown for a random selection of ten OCs and ten MGs in the high-quality sample within 250~pc. Errors on masses are in the brackets. The full table is available in the online material, and includes all parameters derived in Paper~II. 
}
\end{table*}

\paraphrased{We derived Jacobi radii and masses for 6956 clusters that are not globular clusters and that are closer than 15~kpc. Since clusters further than 15~kpc away were not included in the training data for our Paper~II neural network, their age and extinction estimates were too unreliable to be used.} An attempt was made to fit isochrones to cluster photometry using other methods, although most of these distant clusters had poor quality CMDs, making it impossible to derive accurate estimates of their parameters using \emph{Gaia} data alone. These distant clusters should be investigated separately in a different work, particularly since our adopted model of the Milky Way's potential used to calculate $\Omega$ and $k$ may be less accurate at distances greater than 15~kpc. In the next section, we present these overall results and compare our cluster masses with literature values.


\subsection{Updated definitions for clusters from Paper~II}
\label{sec:discussion:definition}

\paraphrased{Thanks to our Jacobi radius inference method, we are now able to provide updated definitions for the clusters in our Paper~II catalogue. In the following section, we discuss how the incorporation of this method changes our catalogue.}



\paraphrased{We calculated the probability that a cluster has a valid Jacobi radius, $P(r_J)$, the distribution of which is shown in Fig.~\ref{fig:results:jacobi_radii_distribution}.} \drafttwo{827 clusters are strongly incompatible with having a bound component, with $P(r_J)<0.05$. On the other hand, around 5733 clusters are strongly compatible with having a valid Jacobi radius ($P(r_J)>0.95$), with 397 clusters having values between these two limits}. \paraphrased{Masses and Jacobi radii appear to be a successful method for differentiating between bound and unbound objects.} Unlike previous attempts to use the virial theorem to discriminate between OCs and MGs during the preparation of this work (see Sect.~\ref{sec:theory}), the probability that a given cluster has a valid Jacobi radius is more successful at distinguishing between bound and unbound clusters.

\paraphrased{Nevertheless, our current method still appears to have limitations at the low-mass end. Some MGs from Paper~II, such as the densest region of the $\beta$~Tucanae MG, are measured as having small Jacobi masses -- typically less than 40~$\mathrm{M}_\sun$, but often lower than 20~$\mathrm{M}_\sun$. While this suggests that these clusters have compact low-mass bound regions that lie somewhere between the definitions of a multiple star system or a star cluster, there are also multiple reasons why these low Jacobi radii may be errors. Firstly, by assuming a Kroupa IMF, our mass estimates will be biased towards conservative, higher values for dynamically evolved stellar groups that have lost low mass stars in long-term two-body interactions. In these cases, a dense group of a dozen high mass stars will have an overestimated total mass using our method, which will be especially the case for MGs that are mass segregated. Secondly, an implicit assumption in our use of Eqn.~\ref{eqn:jacobi_radius} is that a star cluster is spherically symmetric \citep{binney_galactic_dynamics_1987}. This assumption may break down for small groups of a dozen stars in the densest region of an MG. Some examples of low-mass components of MGs that appear to have a valid Jacobi radius clearly violate this assumption, and may hence be erroneously measured as having a valid Jacobi radius.}

\paraphrased{Consequently, we also recommend using an additional minimum $M_J$ of 40~M$_\sun$ when deciding between OCs and MGs. This lower limit is higher than the $M_J$ of even the smallest widely accepted OCs, such as Melotte~111 (Coma Ber) or Platais~9, but excludes edge cases that appear to be dense regions of MGs where our method breaks down, or cases that may be better classified as a resolved multiple star system. Clusters below this mass limit that have a measured high value of $P(r_J)$ would still be interesting objects for a follow-up study on why some MGs appear to have dense cores. Such dense cores could, for instance, be the remnant of a dissolved OC.}

\begin{table}[t]
    \caption{\label{tab:catalogue_members}Member stars of OCs and MGs within 15~kpc with individual stellar masses.}
    \centering
    \begin{tabular}{lccccccccc}
        \hline\hline
        Name & Source ID & In $r_J$ & Mass (M\textsubscript{$\odot$}) \\
        \hline
        \multicolumn{3}{c}{$\cdot \cdot \cdot$} \\ 
        
        Blanco 1 & 2380571935471330560 & 0 & 0.98$^{+0.01}_{-0.01}$ \\
        Blanco 1 & 2320987858469300864 & 1 & 0.90$^{+0.01}_{-0.01}$ \\
        Blanco 1 & 2320786540467288320 & 1 & 0.48$^{+0.01}_{-0.01}$ \\
        Blanco 1 & 2332908729178258560 & 1 & 0.41$^{+0.01}_{-0.01}$ \\
        Blanco 1 & 2320757850084772480 & 1 & 0.23$^{+0.01}_{-0.03}$ \\
        Blanco 1 & 2320550046683235328 & 1 & 1.09$^{+0.01}_{-0.01}$ \\
        Blanco 1 & 2332928451667849856 & 1 & 0.46$^{+0.01}_{-0.01}$ \\
        Blanco 1 & 2314778985026776320 & 1 & 0.37$^{+0.01}_{-0.02}$ \\
        Blanco 1 & 2334068503491984896 & 0 & 0.22$^{+0.02}_{-0.03}$ \\
        Blanco 1 & 2330660983812933376 & 0 & 0.89$^{+0.01}_{-0.01}$ \\
        
        \multicolumn{3}{c}{$\cdot \cdot \cdot$} \\ 
        \hline
    \end{tabular}
    \tablefoot{Shown for ten member stars of Blanco 1. The full table is available in the online material, and includes all parameters listed in the Paper~II table of member stars from \emph{Gaia} DR3. 
}
\end{table}

\begin{table}[t]
\caption{\label{tab:catalogue_results}\paraphrased{Total counts of cluster types.}}
\centering
\begin{tabular}{lcp{22mm}cc}
\hline\hline
Type & Label & Criteria & Count & (high q.)\tablefootmark{a} \\
\hline
OC & \texttt{o} & $P(r_J)\geq0.5$ and\newline$M_J\geq~40$~$\mathrm{M}_\sun$ & 5647 & 3530\\
\hline
MG & \texttt{m} & $P(r_J)~<~0.5$ or\newline$M_J~<~40$~$\mathrm{M}_\sun$ & 1309 & 539 \\
- no $r_J$ & & $P(r_J) < 0.5$ & 992 & 301 \\
- has $r_J$ & & $P(r_J) \geq 0.5$ & 317 & 238 \\
\hline
GC & \texttt{g} & Crossmatching\tablefootmark{b} & 132 & 25 \\
\hline
Too distant & \texttt{d} & $d \geq 15$~kpc & 62 & 6 \\
\hline
Rejected & \texttt{r} & Manual\tablefootmark{c} & 17 & - \\
\hline
\end{tabular}
\tablefoot{
\tablefoottext{a}{\paraphrased{Count of how many clusters of a given type are also in the high-quality sample of clusters from Paper~II, which are those with a median CMD class greater than 0.5 and an astrometric S/N (CST) greater than 5$\sigma$.}}
\tablefoottext{b}{Clusters defined as GCs in \cite{vasiliev_gaia_edr3_2021}, \cite{kharchenko_global_2013}, or \cite{gran_hidden_haystack_2022}.}
\tablefoottext{c}{Clusters later matched to galaxies or dwarf galaxies, or removed due to being obvious clustering algorithm errors (see Sect.~\ref{sec:discussion:definition}).}
}
\end{table} 

\paraphrased{In total, our Paper~II catalogue contains 5647 clusters (82\%) with $P(r_J)>0.5$, a minimum mass of 40~$\mathrm{M}_\sun$, and at least ten observed stars within $r_J$. In the solar neighbourhood, most clusters from Paper~II are classified as MGs, with 11\% (26 of 234) clusters within 250~pc being compatible with our OC definition.} \drafttwo{Within 100~pc, there are only two OCs: Melotte~25 (the Hyades) and Melotte~111 (Coma~Ber).} Of the new clusters reported in Paper~II, 1441 of 2387 are compatible with being OCs, or 487 of the 739 high quality new clusters from Paper~II. This is in line with our belief in Paper~II that a significant fraction of our newly reported clusters did not appear to be OCs. Surprisingly, seven new clusters reported in Paper~II within 250~pc are compatible with being OCs, although all but one have low masses of 66~$\mathrm{M}_\sun$ or lower.

\drafttwo{An updated version of the Paper~II catalogue including object classifications and masses is given in Table~\ref{tab:catalogue_clusters}. In addition, an updated version of the Paper~II stellar membership lists for each cluster (including individual stellar masses) is given in Table~\ref{tab:catalogue_members}.} \paraphrased{Table~\ref{tab:catalogue_results} shows overall statistics on the total number of clusters by object type and sample.}

\drafttwo{In addition, some naming updates were incorporated into the catalogue. Firstly, eleven additional clusters were updated to be labelled as GCs: firstly, HSC~134 and HSC~2890, which are in fact Gran~3 and Gran~4 and were already reported in \cite{gran_hidden_haystack_2022}. In addition, Palomar~2, 6, 8, 10, 11, and 12, IC~1276, 1636-283 (whose name was changed to the more widely used ESO~452-11), and Pismis~26 were updated to be labelled as GCs as in \cite{kharchenko_global_2013} and \cite{perren_unified_cluster_2023}.}

\drafttwo{Next, 17 clusters clearly compatible with galaxies, dwarf galaxies, or errors in our clustering algorithm are flagged in the catalogue ($\mathrm{Type}=\texttt{r}$). These clusters were highlighted by members of the community in the months since the publication of Paper~II (Großschedl, private communication; Alessi, private communication), and should not be used in studies of galactic star clusters.}

\drafttwo{Finally, we incorporated a handful of naming corrections from \cite{perren_unified_cluster_2023}, Teutsch~(private communication), \referee{and Röser and Schilbach~(private communication)}. These corrections are listed in Table~\ref{tab:catalogue_name_changes}, and include corrections to typos, changes to certain names for consistency with other clusters with the same designation, \referee{changes to clusters from \cite{liu_catalog_newly_2019} to have the more common designation `LP' instead of `FoF',} and incorporation of a paper missed from crossmatching in Paper~II and by other input OC catalogues \citep{kronberger_new_galactic_2006}. Clusters retain the same \texttt{id} number between this work and Paper~II, and previous names from Paper~II are listed in the full online catalogue.}

\subsection{Overall catalogue distributions}

\begin{figure*}[t]
    \centering
    \includegraphics[width=\textwidth]{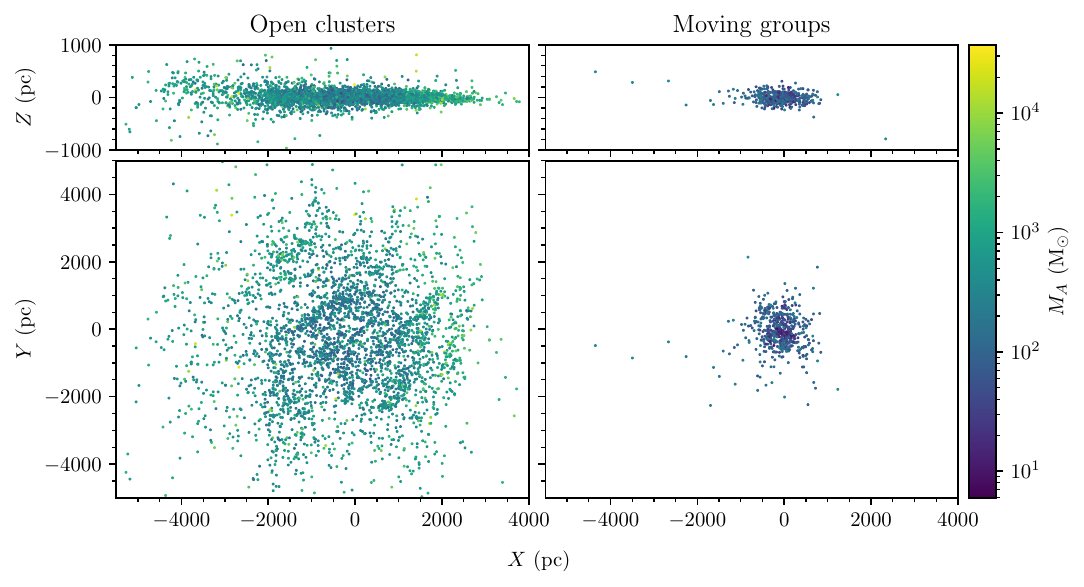}
    \caption{Comparison of the spatial distribution of OCs and MGs. \emph{Left column:} distribution in \referee{Cartesian heliocentric coordinates} of 3530 OCs in the high quality sample of OCs from Table~\ref{tab:catalogue_results}. \referee{The Sun is at $X=Y=0$~pc, the galactic centre is to the right, and the $Z$ axis denotes height above or below the plane.} OCs are shaded by the mass of the entire detected cluster, including tidal tails. \emph{Right column:} identical plot, but for 539 MGs in the high quality sample.}
    \label{fig:discussion:xyz_distribution}
\end{figure*}

\begin{figure}[t]
    \centering
    \includegraphics[width=\columnwidth]{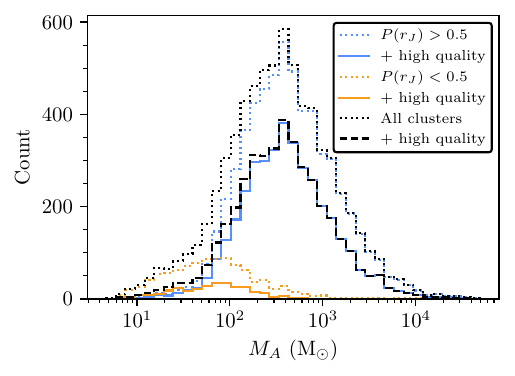}
    \caption{\paraphrased{Histogram of total cluster masses $M_A$ for all clusters divided into different samples. This is shown for all clusters (black dotted line), those with $P(r_J) > 0.5$ (blue dotted line), and those with $P(r_J) < 0.5$ (orange dotted line). The dashed and solid variants of these lines show the mass distribution for these clusters but restricted to only those in the high quality object sample. \citep[Adapted from][]{hunt_improving_census_2023}}}
    \label{fig:results:mass_distribution}
\end{figure}

\paraphrased{In this subsection, we compare differences between the spatial and parameter distributions of OCs and MGs. Figure~\ref{fig:discussion:xyz_distribution} shows the distribution in Cartesian heliocentric coordinates of the high-quality samples of OCs and MGs. In Paper~II, we remarked that our catalogue had a nonphysical peak in density near to the Sun, with many hundreds of additional clusters compared to catalogues such as that of \cite{cantat-gaudin_clusters_mirages_2020}, which we suggested were MGs. Figure~\ref{fig:discussion:xyz_distribution} confirms this hypothesis, showing that objects now classified as MGs were responsible for the density peak near to the Sun, as MGs are all at much lower distances.} As suspected from Paper~II, MGs dominate the distribution of clusters in the catalogue near to the Sun.

\paraphrased{OCs and MGs in this work have different parameter distributions, with one particularly strong difference being in their masses. The distributions of OC and MG masses in Fig.~\ref{fig:results:mass_distribution} show that MGs in this work are generally much less massive, with a modal total mass of $\sim75$~$\mathrm{M}_\sun$. OCs are generally much heavier, with a modal mass of around$\sim400$~$\mathrm{M}_\sun$.}

\begin{figure}[t]
    \centering
    \includegraphics[width=\columnwidth]{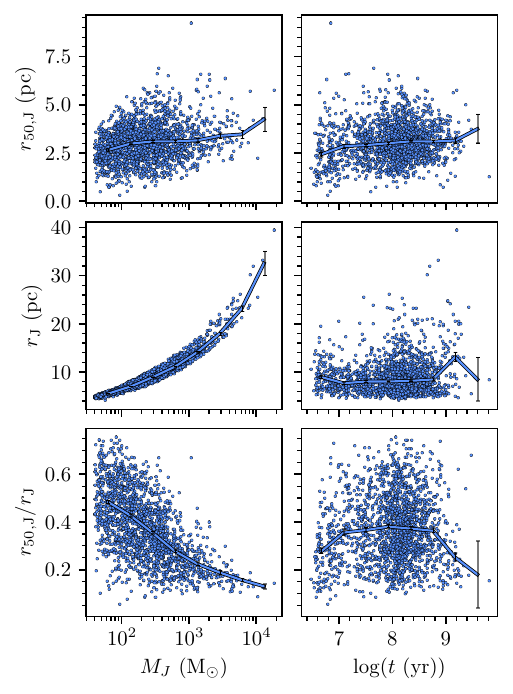}
    \caption{\referee{Jacobi} radii and concentrations of high-quality OCs within 2~kpc, shown for $r_{50,\text{J}}$ (\emph{upper row}), \referee{$r_\text{J}$ (\emph{middle row}),} and cluster concentrations $r_{50,\text{J}}/r_\text{J}$ (\emph{bottom row}) against cluster \referee{Jacobi} mass (\emph{left column}) and cluster age (\emph{right column}). \referee{In each panel, a trend line of binned medians is shown in blue, with error bars showing standard error.}}
    \label{fig:discussion:correlations}
\end{figure}

\begin{figure}[t]
    \centering
    \includegraphics[width=\columnwidth]{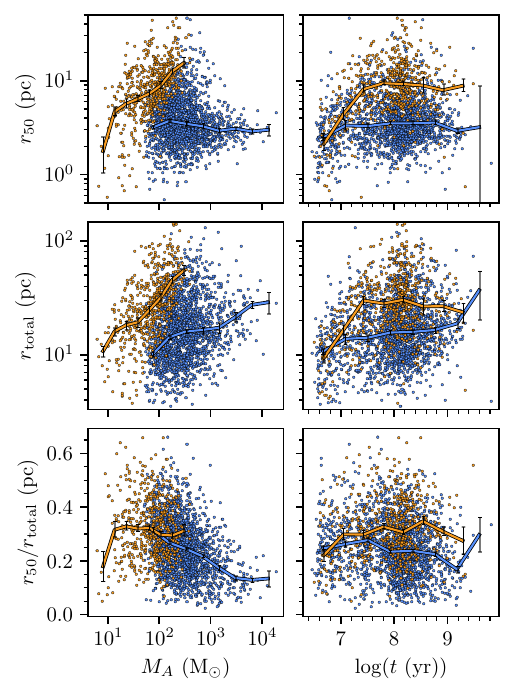}
    \caption{Radii and concentrations of \referee{high-quality OCs (blue) and high-quality MGs (orange)} within 2~kpc, shown for $r_{50}$ (\emph{upper row}), \referee{$r_\text{total}$ (\emph{middle row}),} and cluster concentrations $r_{50}/r_\text{total}$ (\emph{bottom row}) against \referee{total} cluster mass (\emph{left column}) and cluster age (\emph{right column}). \referee{Trend lines are plotted with the same formatting as in Fig.~\ref{fig:discussion:correlations}.}}
    \label{fig:discussion:correlations_mgs}
\end{figure}

OC and MG radii also show a number of interesting differences and correlations that can be compared against theoretical predictions. Figure~\ref{fig:discussion:correlations} shows OC radii and concentrations against mass and age: namely, the radius containing 50\% of members within $r_J$ ($r_{50,\text{J}}$), \referee{$r_J$ itself,} and the ratio between these two radii (analogous to the concentration of the cluster) $r_{50,\text{J}}/r_J$. These are only shown for OCs in the high-quality sample of objects that are within 2~kpc; namely, those for which radii, masses and ages are most robustly measured. As a function of mass, \referee{$r_{50,\text{J}}$} is lightly correlated, with higher mass clusters generally having slightly larger cores. \referee{$r_\text{J}$ is strongly correlated with mass -- although this is to be expected, since $r_\text{J}$ is calculated directly from cluster mass with Eqn.~\ref{eqn:jacobi_radius}.} Cluster concentrations are also strongly correlated with cluster mass, with the lowest mass clusters being the least centrally concentrated, strongly suggesting that OCs are less centrally concentrated as a function of mass, likely due to dynamical processes within them \citep{portegies_zwart_young_2010,krause_physics_2020}. However, as a function of age, cluster radii and concentrations are generally uncorrelated, although the youngest clusters ($\log t < 7$) may be slightly smaller and more concentrated, which is in line with existing theory that OCs undergo a phase of expansion \citep{krause_physics_2020}. Since the minimum age that the neural network from Paper~II can measure is $\log t = 6.4$, young cluster ages may not be well measured enough to adequately sample this range of cluster formation.

Although our methodology was not originally intended to detect MGs (Paper~I), the MGs in our catalogue are still an interesting point of comparison against our detected OCs. Figure~\ref{fig:discussion:correlations_mgs} shows cluster median radii $r_{50}$, total radii including all member stars and any tidal tails $r_\text{total}$, and the ratio between cluster median radius over total radius $r_{50}/r_\text{total}$ for the \referee{high-quality OCs and MGs in our sample within 2~kpc.} The size of detected MGs strongly correlates with their mass -- although this may be a selection effect, as it could be easier to detect member stars of MGs out to higher radii on-sky if they are also higher mass. \referee{MGs clearly occupy a different region of radius-mass parameter space, generally being much larger than OCs at a given mass.} MG concentration does not appear to change as a function of mass, which is different to OCs whose structural evolution is driven by their internal (bound) dynamics and gradual dissolution due to the Milky Way's potential \citep{krause_physics_2020}. 

\referee{MGs and OCs differ strongly as a function of age. OCs and MGs have similar sizes at young ages for $\log t < 7$, suggesting a similar origin. However, whereas OCs only undergo a small phase of expansion, MGs expand much more strongly, eventually being significantly larger than OCs at all older ages (particularly for $r_{50}$.)} This increase in observed size is consistent with the MGs in our catalogue being unbound groups of coeval stars that expand over time. Nevertheless, many MGs are older than the expected time it would take for them to disperse \citep{zucker_disconnecting_dots_2022}. If these MGs are real, co-evolutionary groups of stars, then they could be unbound remnants of bound OCs. These objects (and their dynamical evolution, such as whether or not their member stars are expanding from a common origin) should be investigated further in a future work.

\subsection{Comparison of masses with literature results}
\label{sec:results:masses}

\begin{figure*}[t]
    \centering
    \includegraphics[width=\textwidth]{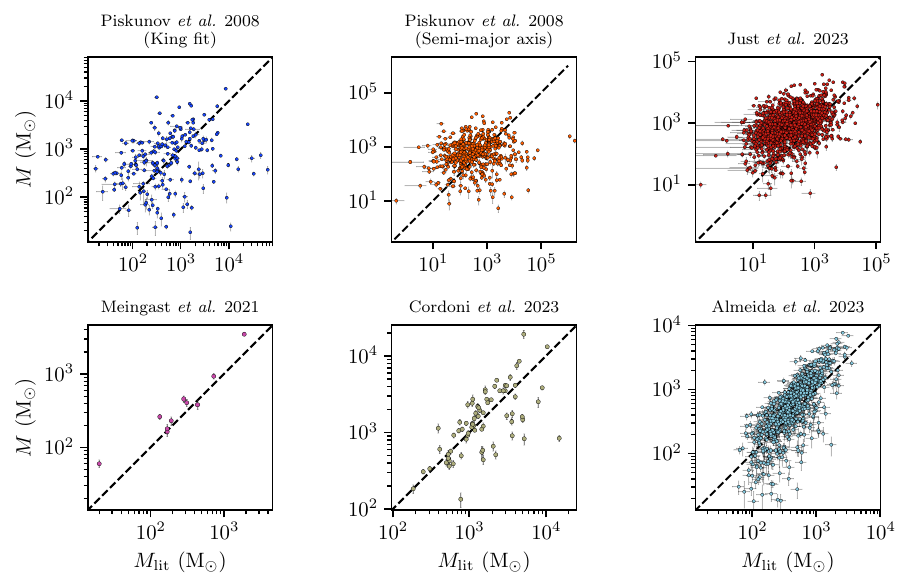}
    \caption{\paraphrased{Cluster masses in this work compared against those in the literature. The $x$-axes show literature mass values while the $y$-axes show cluster Jacobi masses derived in this work. The dashed $y=x$ line shows where mass measurements that are in perfect agreement would be. \citep[Adapted from][]{hunt_improving_census_2023}}}
    \label{fig:results:mass_comparison}
\end{figure*}

Finally, an important step in validating our results is to compare our derived cluster masses against results from the literature, although cluster masses are generally not frequently measured in the literature and different methodologies can produce highly different results. \paraphrased{Cluster masses from the literature are compared against masses derived in this work in Fig.~\ref{fig:results:mass_comparison}. We compare our masses to masses derived without \emph{Gaia} data and using profile fitting techniques in \cite{piskunov_tidal_radii_2008} and \cite{just_global_survey_2023}; using \emph{Gaia} DR2 photometry in \cite{meingast_extended_2021}; and using \emph{Gaia} DR3 photometry in \cite{cordoni_photometric_binaries_2023} and \cite{almeida_revisiting_mass_2023}.}

\paraphrased{We find that our masses are most similar to the small sample of ten nearby clusters studied in \cite{meingast_extended_2021}, who applied a similar methodology of assuming a Kroupa IMF and fitting it to a cluster's mass function. Our mass estimates are higher than theirs, with this discrepancy being largest for Platais~9 -- a cluster that \cite{meingast_extended_2021} estimate to have a mass of only 13.1~$\mathrm{M}_\sun$, compared to our measurement of $62.1 \pm 7.9$~$\mathrm{M}_\sun$. In fact, our mass estimates are generally higher than the estimates of all \emph{Gaia}-based works. This is likely due to our incorporation of corrections for selection effects and unresolved binaries, both of which will cause our mass estimates to be higher than existing \emph{Gaia} works that do not correct for both effects. Even amongst \emph{Gaia}-based works, there is currently little general agreement on the masses of most clusters.}

\paraphrased{We have limited similarity in mass measurements for some clusters to the OC mass catalogue of \cite{cordoni_photometric_binaries_2023}, who fitted bespoke mass functions to clusters in their sample but without correcting for incompleteness. Some clusters in their work have significantly higher cluster masses than in this work, which is likely due to the bespoke cluster mass functions that they use. The cluster with the largest discrepancy is Haffner~26, which we measure to have a mass of $868\pm86$~$\mathrm{M}_\sun$, compared to their mass of 14563~$\mathrm{M}_\sun$. For Haffner~26, \cite{cordoni_photometric_binaries_2023}'s fitted mass function has power law indices of 3.37 and 4.78 above and below a break point at 1~$\mathrm{M}_\sun$. This mass function is much steeper than the Kroupa IMF used in this work, which has indices of 2.3 and 1.3 above and below a 0.5~M$_\sun$ break point. However, after correcting for selection effects, our mass function for Haffner~26 is highly compatible with a Kroupa IMF, and is strongly incompatible with the strong power law indices fitted in \cite{cordoni_photometric_binaries_2023}. In addition, since Haffner~26 is at a distance of around 3~kpc from the Sun, few of its low-mass stars are resolved by \emph{Gaia}. Our approach of assuming a Kroupa IMF may be less accurate for some nearby clusters for which their mass function can be clearly resolved, but it is at least a safe and consistent approach for clusters at all distances. Extrapolation of a steep mass function of 4.78 below 1~$\mathrm{M}_\sun$ in \cite{cordoni_photometric_binaries_2023} likely contributes most of the mass towards this cluster in their measurement, even though few stars below that mass are actually observed by \emph{Gaia} for Haffner~26.}

\rewrite{Finally, \cite{almeida_revisiting_mass_2023} publish a catalogue of cluster masses based on \emph{Gaia} DR3 data, created by extracting estimated stellar masses (including accounting for binaries) through comparison with simulated clusters, then fitting bespoke mass functions to each cluster. The overall trend of our results matches theirs, although our mass estimates are once again generally higher, which is likely due to our additional corrections for \emph{Gaia} incompleteness. Similar to \cite{cordoni_photometric_binaries_2023}, some of our mass estimates for clusters are significantly lower than theirs -- which once again may be due to differences from extrapolating mass functions derived for high mass stars across the entire (unobserved) mass range of a cluster, or due to differences in cluster membership list.}

\paraphrased{Our results show poor agreement with masses derived in pre-\emph{Gaia} works. \cite{piskunov_tidal_radii_2008} presented the largest catalogue of cluster masses that was available before the release of \emph{Gaia}. Their masses are calculated in two ways: firstly, by fitting a \cite{king_structure_star_1962} profile to clusters and assuming that the King tidal radius $r_t=r_J$, then inverting Eqn.~\ref{eqn:jacobi_radius} to derive a cluster mass given its radius; and secondly, by fitting only semi-major axes to clusters and deriving a mass in the same way. However, these methods are extremely sensitive to the derived cluster membership list and cluster radius, since $M_J \propto r_J^3$ in Eqn.~\ref{eqn:jacobi_radius}. Particularly as \cite{piskunov_tidal_radii_2008} relied on cluster membership lists that do not use \emph{Gaia} astrometric data and are hence much harder to clean of field stars, in addition to being less complete \citep{cantat-gaudin_milky_2022}, differences in cluster membership alone can explain why our mass measurements have poor agreement with theirs. Cluster membership lists in \cite{kharchenko_global_2013} contain four times fewer member stars than in our Paper~II catalogue, and even though our approximate King tidal radii in Paper~II were only $\approx 1.5\times$ larger than the tidal radii derived in \cite{piskunov_tidal_radii_2008}, this already corresponds to a $\sim 3\times$ increase in cluster mass based on Eqn.~\ref{eqn:jacobi_radius}. It is hence likely that data and methodological differences alone can explain the large inconsistencies between \emph{Gaia} and pre-\emph{Gaia} cluster masses.} \cite{just_global_survey_2023} also use a similar method relying on pre-\emph{Gaia} OC membership lists, for whom we also have poor agreement with their results. 

\rewrite{In summary, some of our mass results are in good agreement with literature catalogues, although the majority are not. This can be explained by differences in methodology, in particular differences in accounting for selection effects meaning our mass estimates are generally higher, in addition to differences in adopted mass functions and cluster membership lists. OC masses derived with pre-\emph{Gaia} cluster membership lists are generally in poor agreement with masses using \emph{Gaia} data.}

\section{Discussion}
\label{sec:discussion}

\rewrite{To the best of the authors' knowledge, this work represents the largest catalogue of Milky Way star cluster masses ever derived, in addition to being the first to classify clusters robustly into bound and unbound objects. In this section, we discuss a number of interesting scientific use cases of this work, beginning with deriving a mass-dependent completeness estimate of our catalogue.}

\subsection{Completeness of the \emph{Gaia} DR3 open cluster census}
\label{sec:discussion:completeness}

\begin{figure}[t]
    \centering
    \includegraphics[width=\columnwidth]{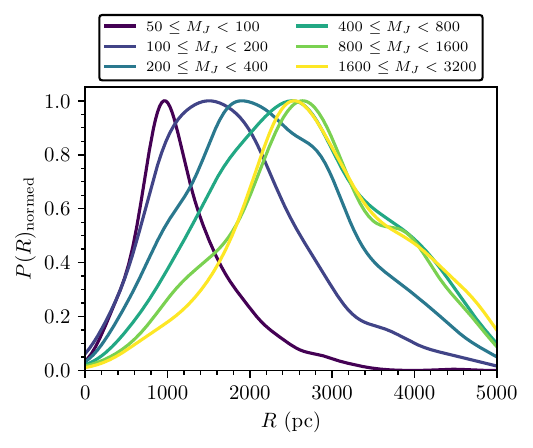}
    \caption{\referee{Kernel density estimates of the two-dimensional distance from the Sun $R=\sqrt{X^2+Y^2}$ distribution of clusters in different mass ranges.} \paraphrased{All curves are normalised to have a peak of one for easier comparison between curves. \citep[Adapted from][]{hunt_improving_census_2023}}}
    \label{fig:discussion:completeness}
\end{figure}

\begin{figure}[t]
    \centering
    \includegraphics[width=\columnwidth]{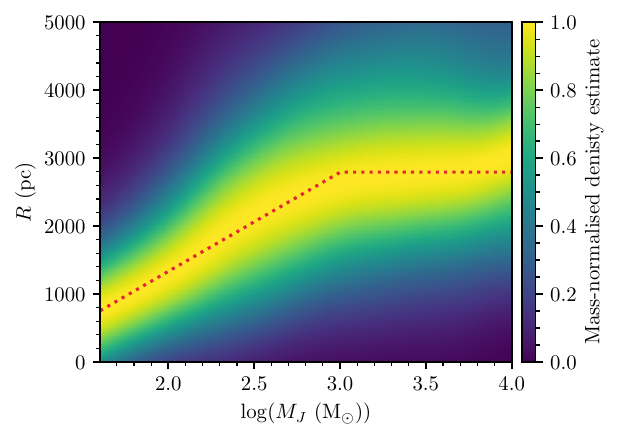}
    \caption{\paraphrased{Kernel density estimate of the $R$-mass distribution of clusters in this work. To enhance the clarity of the peak of this distribution, the density estimate at each mass is normalised to have a peak of one. The best-fit log-linear completeness model (see Sect.~\ref{sec:discussion:completeness}) is shown by the red dotted line. \citep[Adapted from][]{hunt_improving_census_2023}}}
    \label{fig:discussion:completeness_kde}
\end{figure}

The completeness of the OC census is an important but difficult to measure quantity. \paraphrased{For instance, although \cite{kharchenko_global_2013} derived that their OC catalogue was complete within 1.8~kpc, this claim has since been disproven by many studies that report new OCs within this distance using \emph{Gaia} data \citep[e.g.][]{castro-ginard_new_2018,castro-ginard_hunting_2019,castro-ginard_hunting_open_2020,castro-ginard_hunting_open_2022,liu_catalog_newly_2019,sim_207_2019,hunt_improving_2021,hunt_improving_open_2023}}. Any investigation of the OC census must be conducted carefully. \paraphrased{In the Gaia era, \cite{anders_milky_way_2021} derive a completeness estimate of the OC census, although this was performed without cluster masses, with it being unknown how masses may affect the completeness of an OC census.} In this section, using our catalogue of cluster masses, we will derive an approximate mass-dependent completeness estimate for our catalogue, demonstrating the importance of cluster masses in deriving the completeness of the OC census.

It is helpful to first consider what the distribution of OCs as a function of radius from the Sun should be. Since the scale height of OCs in the disk is small ($\sim100$~pc) compared to the kpc-scales out to which OCs are observed, one can approximate the expected OC distribution in two dimensions $X$ and $Y$ looking at the galaxy top-down. Given a uniform top-down surface density of clusters per square parsec $n$, the expected number of clusters $N$ within some radius $R=\sqrt{X^2+Y^2}$ is hence given by:

\begin{equation}
    N(R) = n \pi R^2,
    \label{eqn:completeness}
\end{equation}

\noindent with a derivative of:

\begin{equation}
    \frac{\text{d}N(R)}{\text{d}R} = 2n \pi R,
    \label{eqn:completeness-derivative}
\end{equation}

\noindent implying that the radius distribution of the OC census should increase linearly, assuming that $n$ is constant and that $R$ does not exceed the distance to the edge of the Milky Way's disk. 

In practice, the actual observed distribution of OCs is unlikely to follow this simple model exactly. Estimation of the true completeness of the OC census is challenging, as the distribution of OCs depends on some distribution function of OCs in the Milky Way, and cannot be assumed to be uniform \citep{anders_milky_way_2021}. For instance, the distribution of young OCs is known to be correlated with the Milky Way's spiral arms \citep{castro-ginard_milky_way_2021}. Deriving such a model for OCs is beyond the scope of this work; however, we can produce a rough estimate of the OC completeness distribution as a function of mass as a proof of concept.

\paraphrased{As an initial test, the $R$ distribution of clusters when divided into separate mass bins in Fig.~\ref{fig:discussion:completeness} shows clear signs of incompleteness depending on cluster mass, with lower mass clusters being more likely to occur at low distances in our catalogue. In the lowest mass bin ($50 \leq M_J < 100$~$\mathrm{M}_\sun$), the cluster distribution peaks at $\sim 1$~kpc, while it peaks at $\sim 2.7$~kpc for the two highest mass bins ($800 \leq M_J < 1600$~$\mathrm{M}_\sun$ and $1600 \leq M_J < 3200$~$\mathrm{M}_\sun$).} In the four lowest mass bins, $P(R)_\text{normed}$ appears roughly linear up to a peak, after which the distribution falls off exponentially. This is roughly the expected model of the OC distribution implied by Eqns.~\ref{eqn:completeness}~and~\ref{eqn:completeness-derivative}, given some limiting 100\% completeness radius $R_{100\%}$ at each given mass. On the other hand, the highest mass bins do not appear linear up to their peak radius. This may be because high-mass clusters seem more likely to be found in the direction of the galactic centre (see Fig.~\ref{fig:discussion:xyz_distribution}), and that assuming their $n$ is uniform is a poor assumption.

To investigate the distribution of clusters without mass binning, Fig.~\ref{fig:discussion:completeness_kde} shows the complete mass-$R$ distribution of OCs smoothed with kernel density estimation. \paraphrased{Kernel density estimates were normalised based on mass, in effect meaning that every vertical strip in the figure has a peak at one, helping to make clear where the distribution peaks at a given cluster mass. The trend in peaks shows a log-linear relation up to a mass of $\sim 1000$~M$_\sun$, after which the distribution does not rise further. This suggests that $\sim 2800$~pc is the approximate upper limit of \emph{Gaia}'s 100\% completeness.} \drafttwo{This could be due to multiple limitations of current \emph{Gaia} DR3 data, such as its magnitude limit, astrometric accuracy, and extinction.}

\paraphrased{To quantify this relationship, we fitted a log-linear model with a break point after which the model is flat to the peaks of this distribution from $40 \leq M_J \leq 10^4~\text{M}_\sun$. This gives the approximate 100\% completeness limit of our OC census $R_{100\%}$, with the model taking the form:}

\begin{equation}
    R_{100\%} = 
    \begin{cases}
    \alpha \log ( M_J~[M_\sun] ) + \beta & R_{100\%} < R_\text{break} \\
    R_\text{break} & R_{100\%} \geq R_\text{break}
    \end{cases}
\end{equation}

\noindent \paraphrased{where the constraint $0 \geq R_{100\%} < R_\text{break}$ was also applied during fitting.} \referee{Our best fit had values $\alpha=633.1\pm7.3$~pc, $\beta=-1582.6\pm39.5$~pc, and $R_\text{break}=2792.9\pm8.2$~pc.}

\paraphrased{This model is at clear odds with the claim of \cite{kharchenko_global_2013}, who claimed that the OC census is complete within 1.8~kpc. Within 1.8~kpc, our all-sky OC census is only complete for clusters heavier than $\sim230$~M$_\sun$ -- a cluster mass similar to that of Melotte~25 (the Hyades). Our catalogue is only complete within 1~kpc for clusters heavier than 100~$\mathrm{M}_\sun$.}


\subsection{Estimating the age and mass functions of OCs in the Milky Way}
\label{sec:discussion:age_mass_function}

\begin{figure}[t]
    \centering
    \includegraphics[width=\columnwidth]{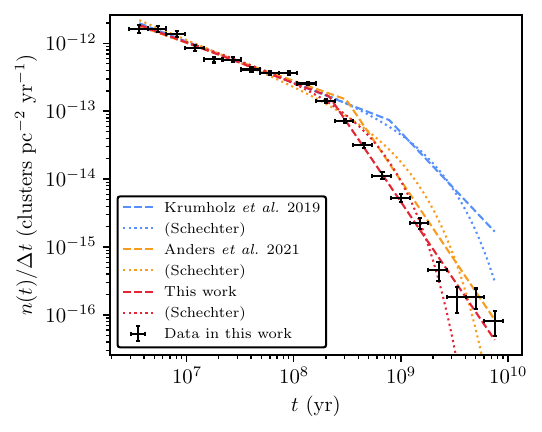}
    \caption{Completeness-corrected age function of OCs in this work (black points) compared against various age functions in the literature. Dashed lines show broken power law fits while dotted lines show Schechter function fits. Literature age functions are normalised to ages below 0.2~Gyr for easier visualisation of differences in shape of the upper end of the distributions. The blue lines show fits from \cite{krumholz_star_2019}, the orange lines from \cite{anders_milky_way_2021}, and the red lines are fits from this work. Poisson uncertainties on the data are indicated by the error bars. \changed{Schechter function fits from \cite{krumholz_star_2019} and \cite{anders_milky_way_2021} have characteristic ages scaled by a factor $\log_{10}{\mathrm{e}}$ to correct for an error in their Schechter function fitting codes.}}
    \label{fig:discussion:age_number_density}
\end{figure}

\begin{figure}[t]
    \centering
    \includegraphics[width=\columnwidth]{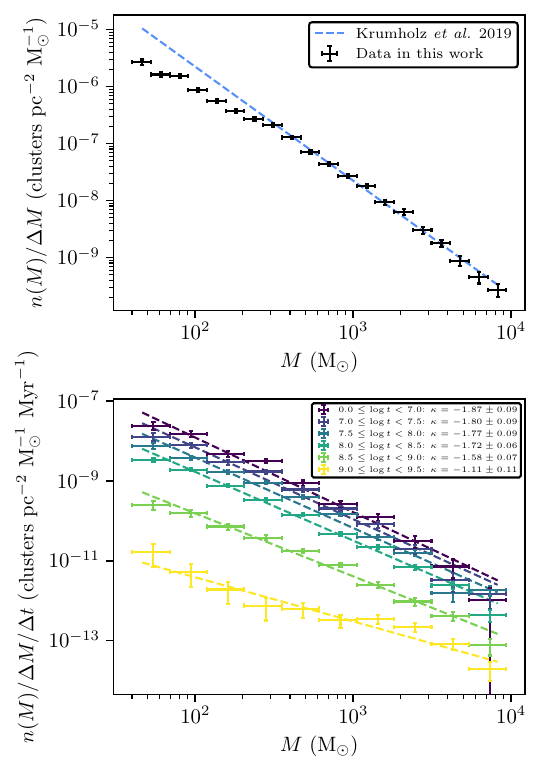}
    \caption{Mass function of OCs in this work. \emph{Top:} Completeness-corrected mass function of OCs in this work (black points) compared against a $\kappa=-2$ power law \citep[][blue dashed line]{krumholz_star_2019} fit to clusters with masses greater than 400~$\text{M}_\sun$. \emph{Bottom:} completeness-corrected mass functions for clusters, separated into age ranges and including power-law fits to each age range.}
    \label{fig:discussion:number_density}
\end{figure}

\begin{figure}[t]
    \centering
    \includegraphics[width=\columnwidth]{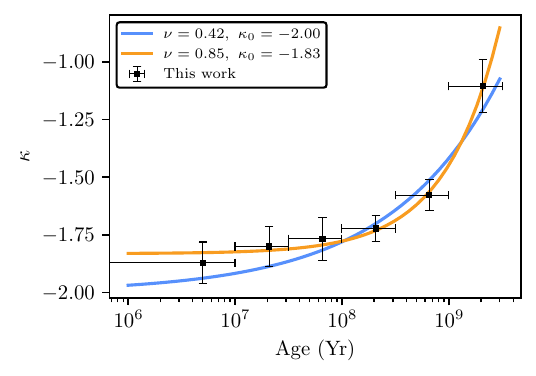}
    \caption{Slope of power-law fits in Fig.~\ref{fig:discussion:number_density} as a function of age. A fit to this slope with $\kappa_0$ fixed to -2 is shown in blue, with a fit with $\kappa_0$ free shown in orange.}
    \label{fig:discussion:number_density_slope}
\end{figure}

\begin{figure}[t]
    \centering
    \includegraphics[width=\columnwidth]{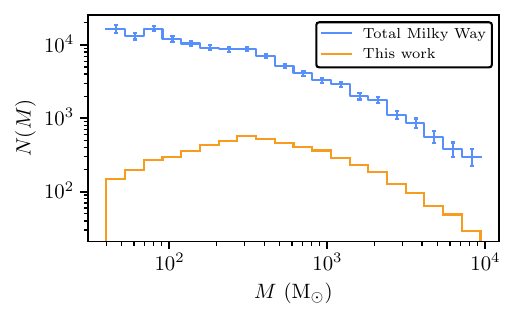}
    \caption{Completeness-corrected estimated total number of OCs in the Milky Way as a function of mass ($N(m)$, blue) including Poisson uncertainties on bins, compared against the distribution of OC masses in this work (orange).}
    \label{fig:discussion:total_number}
\end{figure}

Using the approximate completeness estimate in Sect.~\ref{sec:discussion:completeness}, it is also possible to estimate the age and mass functions of OCs in the Milky Way from the number density of OCs as a function of age or mass, in addition to the total number of OCs in the Milky Way. To do so, the number of OCs at \drafttwo{distances} below $R_{100\%}$ at their given mass was counted into bins, and then divided by the total 2D area of a circle of radius $R_{100\%}$ at the central mass of each bin. 

Our completeness-corrected age function for OCs in the Milky Way is plotted in Fig.~\ref{fig:discussion:age_number_density}. Unlike \cite{krumholz_star_2019} and \cite{anders_milky_way_2021}, who find that the cluster age function is well approximated by a broken power law or a Schechter function, we find that our cluster age function is only fitted well by a broken power law, with a much sharper `knee' in our cluster age function than that of \cite{anders_milky_way_2021} or \cite{krumholz_star_2019}. This may be due to our different definition of an OC in terms of its gravitational potential, which may mean our catalogue is more strongly cleaned of older, unbound star clusters. Nevertheless, we confirm the results of \cite{anders_milky_way_2021}, who found that the number of old clusters in \emph{Gaia} is much lower than previous pre-\emph{Gaia} results such as \cite{piskunov_global_2018}. Based on our results in Paper~II, it is likely that the reduced number of old OCs in \emph{Gaia}-derived results (including this work) is due to many old OCs reported before \emph{Gaia} being unlikely to be real. Our broken power law fit to our data gives $\alpha_1= -0.594 \pm 0.038$, $\alpha_2= -2.321 \pm 0.127$, and with a break point at $\log t_\text{break}= 8.33 \pm 0.04~\log\,\text{yr}$. The slopes of this distribution are compatible within uncertainty to the results of \cite{anders_milky_way_2021}, although our $\log t_\text{break}$ is slightly lower than their value of $\log t_\text{break} = 8.49 ^{+0.21} _{-0.21}~\log\,\text{yr}$.

The upper plot of Fig.~\ref{fig:discussion:number_density} shows our completeness-corrected mass function for OCs in the Milky Way. Above a mass of around 400~$\text{M}_\sun$, this mass function is well approximated by a power law with index $\kappa=-2$, which is identical to the cluster initial mass function found in numerous other galaxies that is well approximated by a power law with slope $\kappa=-2$ for clusters with masses below $\approx 10^4~\text{M}_\sun$ \citep{portegies_zwart_young_2010, krumholz_star_2019}, implying a log-uniform rate of cluster formation as a function of mass. However, for clusters below a mass of 400~$\text{M}_\sun$, we find a slightly less steep power law function is a better fit to the data. \drafttwo{This appears to be a trend based on age. The lower plot of Fig.~\ref{fig:discussion:number_density} shows the age-binned mass function of the same clusters, including fits by unbroken power laws. For the youngest clusters, their mass function is close to a $\kappa=-2$ power law, which is the prediction for young clusters in \cite{krumholz_star_2019}. With increasing age, the cluster mass function appears to flatten, in addition to decreasing at all masses. This flattening of the mass function slopes with age may suggest accelerated cluster dissolution for low-mass clusters as a function of age compared to high-mass clusters, and should be investigated with theoretical studies.} 

\drafttwo{Figure~\ref{fig:discussion:number_density_slope} shows the slopes of power law fits to the age-binned mass function of clusters as a function of age. We fit offset power laws of the form $\kappa=At^\nu + \kappa_0$ to this distribution; firstly, for $\kappa_0$ fixed to -2, as predicted for zero-age clusters \citep{krumholz_star_2019}; and secondly, with all parameters free. In the first ($\kappa_0$ fixed) case, we find $\nu=0.423\pm0.078$; in the second case, we find $\kappa_0=-1.832\pm0.024$ and $\nu=0.853\pm0.111$. These observations should be compared to large-scale N-body simulations of cluster dissolution in the future.}

We also used our derived cluster number density to calculate an estimate of the total number of OCs in the Milky Way at a given mass, $N(M)$, assuming a flat disk distribution of OCs with a radius of 12.5~kpc, which corresponds to the approximate limit out to which OCs are observed in the galactic disk (see e.g. Fig.~\ref{fig:discussion:xyz_distribution}). Figure~\ref{fig:discussion:total_number} shows the total number of OCs in this work compared against a completeness-corrected estimated total number of OCs in the Milky Way. Summing this distribution, we estimate that the Milky Way contains a total of $\sim1.3 \times 10^5$ OCs with masses in the range $40 < M_\text{J} < 10^4$~$\text{M}_\sun$, which is comparable to the $\sim 10^5$ OCs in the Milky Way estimated to exist by \cite{dias_new_catalogue_2002}. This estimated total number implies that only around $\sim 4\%$ of the Milky Way's total number of OCs are known at this time, with this incompleteness being strongest for low-mass clusters. 

\drafttwo{Summing our predicted $N(M)$ distribution, we estimate that the Milky Way contains $\sim4.8 \times 10^7$~$\text{M}_\sun$ of stars that are currently bound to OCs. \cite{cautun_milky_way_2020} used \emph{Gaia} DR2 to estimate that the Milky Way contains $5.04^{+0.43}_{-0.52} \times 10^{10}$~M\textsubscript{$\sun$} of stellar mass; compared to our prediction, this suggests that around 0.1\% of the Milky Way's stars are currently in an OC. This is similar to the ratio between our total number of input stars in Paper~II and the final number of stars that we find to be currently bound to an OC. In Paper~II, we used an input list of 729~million stars from \emph{Gaia} DR3 to construct our catalogue. In this work, we find that 614\,358 of those stars are currently within the Jacobi radius of an OC, which is around 0.1\% of the stars considered in our Paper~II clustering analysis.}

\subsection{Comparison between cluster mass functions and the Kroupa IMF}
\label{sec:discussion:kroupa_imf}

\begin{figure*}[t]
    \centering
    \includegraphics[width=\textwidth]{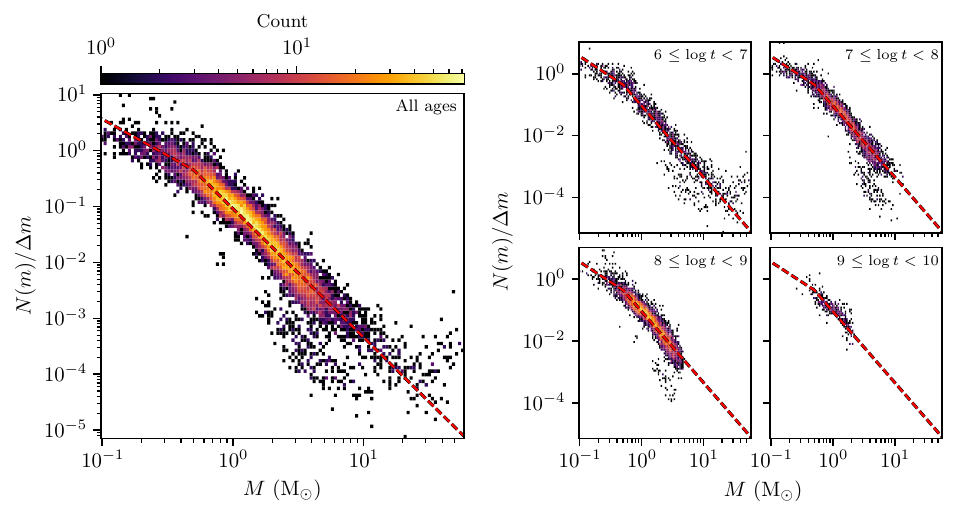}
    \caption{Comparison between mass function points of clusters in this work and the Kroupa IMF. \referee{\emph{Left:}} 2D histograms of all points from all cluster mass functions in this work for 1235 OCs within 2~kpc in the high-quality sample of clusters and with at least 50 member stars compared against the Kroupa IMF (dashed red line). Individual cluster mass functions are normalised before combining. The colour of histogram bins denotes how many mass function points went into each individual bin, corresponding to the colour bar in the upper-right. \referee{\emph{Right:} same as left panel, except clusters are divided into four separate age ranges.}}
    \label{fig:discussion:mfs}
\end{figure*}

Throughout this work, we relied on the Kroupa IMF to calculate total cluster masses. After extensive correction for cluster membership selection effects in Sect.~\ref{sec:masses:selection}, we find that cluster mass functions were widely compatible with the Kroupa IMF, and across a wide range of cluster ages. Figure~\ref{fig:discussion:mfs} shows data points from all mass functions in this work and plotted as a 2D histogram. 

There are some notable outliers in this figure that are worth discussing initially. Firstly, the highest mass points (with masses greater than $\sim 20~\text{M}_\sun$) appear to be over-counted. This is because the cluster ages from Paper~II have a lower limit of $\log t = 6.4$, meaning that high mass stars in star clusters younger than this age cannot have masses higher than this limit assigned to them, and the highest mass bins in young clusters are hence overestimated due to contamination from even higher mass and short-lived O stars. This is visible in the $6 \leq \log t < 7$ subplot of the figure -- only young clusters have these erroneously high measurements. 

Secondly, some mass bins in the range $2 \leq M < 10~\text{M}_\sun$ for some clusters contain around an order of magnitude fewer stars than would be expected for these clusters. These low-count bins have correspondingly high Poisson uncertainties, and do not dramatically change our overall total cluster mass measurements, \referee{but are nevertheless still worth discussion. There are likely to be multiple reasons for the missing high-mass stars in these clusters, including poor-quality CMDs, small number statistics, poor isochrone fits, and unaccounted for selection effects. 902 of the 6956 (12.9\%) of clusters with mass measurements in this work have at least one mass bin more than $5\times$ below the expected value from a Kroupa IMF, and hence have points within the previously identified region. 200 of these clusters have low-quality CMDs (Paper~II CMD score below 0.5) which may mean they are not a real single population of stars or that they are a poor detection of a real cluster, which hence may have gaps for non-physical reasons.} 

\referee{Of the remaining 702 clusters with good-quality CMDs, 572 have fewer than 100 member stars, which could plausibly have gaps simply due to missing stars due to small number statistics, or due to a small number of stars making it difficult to constrain an accurate isochrone fit, as our Paper~II parameter inference accuracy is strongly correlated with number of member stars. Almost all erroneously low mass function points are at the tip of the main sequence within clusters -- a region within a cluster CMD that is generally sparsely populated but that covers a wide range in stellar masses, particularly for young clusters -- meaning that a small error in an isochrone fit or an isochrone itself can correspond to a large error in derived stellar mass, hence faking the appearance of a gap in a measured cluster mass function.}

\referee{Nevertheless, 54 clusters with good quality CMDs and at least 200 member stars still have mass function gaps. Most of these clusters are nearby ($d < 1$~kpc), young ($\log t < 8$), and have well-inferred photometric parameters. All but one of the gaps in these clusters are brighter than $G=12$, are at or near the tip of the main sequence, and occur in well-studied clusters such as Blanco~1 ($9 < G < 9.5$, see Fig.~\ref{fig:masses:selection_effects}). In the case of Blanco~1, this gap appears robust between different works, appearing in other \emph{Gaia}-based works \citep[e.g.][]{zhang_diagnosing_stellar_2020,cantat-gaudin_painting_2020}. The gaps being at brighter magnitudes may be significant for three reasons. Firstly, \emph{Gaia}'s CCDs become saturated above $G=12$, and sources above this limit undergo different photometric processing \citep{riello_gaia_early_2021}. Bright sources in \emph{Gaia} often have much higher astrometric errors, which could mean that they are missed from a cluster membership list due to poor astrometry, that they may only have a two-parameter astrometric solution and were hence not included in our clustering analysis, or that they are more likely to be tagged as a false positive by the \cite{rybizki_classifier_spurious_2022} method we used to clean the \emph{Gaia} DR3 dataset in Paper~II. Secondly, at high magnitudes, there are significantly fewer stars, meaning that \emph{Gaia}'s selection function is much more difficult to accurately characterise empirically. This could impact the \emph{Gaia} or subsample selection functions applied in Sect.~\ref{sec:masses:selection}. With just a handful of stars in a given wide magnitude range to use to empirically determine a selection function, uncertainty on a selection function in a given range is higher. It is notable that in Fig.~\ref{fig:masses:selection_effects}, Blanco~1's subsample selection function is lower in the range where it has a gap, suggesting that one reason for missing stars in this range could be an underestimated selection effect. In the future, it may be necessary to improve subsample selection functions further to be more accurate at bright magnitudes where the presence of few bright stars in most fields makes it difficult to empirically determine subsample selection functions for bright stars accurately. Finally, since stars in the mass range $2 \leq M < 10~\text{M}_\sun$ are usually binaries \citep{moe_mind_your_2017}, it may also be that binary star-induced astrometric errors cause some stars to be missing from our cluster membership lists -- particularly if they are on $\sim 1 $~year orbits with motion similar to that of parallax \citep{lindegren_gaia_early_2021a}. Improved binary star astrometry and classifications in \emph{Gaia} DR4 will help to reduce the number of missed binary stars in future works \citep{gaiacollaboration_gaia_data_2022}.}


Aside from these outliers, the bulk of points in cluster mass functions are a good fit to a Kroupa IMF. Some deviation from a Kroupa IMF is visible for the oldest clusters in Fig.~\ref{fig:discussion:mfs}, where mass functions appear flatter, with a possible physical cause being preferential mass loss of low-mass stars in the oldest clusters. Fundamentally, however, we are unable to reproduce the results of works including \cite{cordoni_photometric_binaries_2023}, who find that cluster mass functions are compatible with power law slopes that deviate significantly from a Kroupa~IMF. To investigate cluster mass functions further in the future, and with a higher accuracy than was possible in this work, cluster mass functions incorporating more accurate cluster-by-cluster binary star corrections should be conducted. This is likely to be possible with future surveys such as \emph{Gaia} DR4, which will provide epoch astrometry for better identification of binaries \citep{gaiacollaboration_gaia_data_2022}, or using stellar spectra in upcoming large spectral surveys such as 4MOST \citep{dejong_4most_4metre_2012} to identify spectroscopic binaries. Currently, photometric identification of binaries with \emph{Gaia} data is only able to detect the highest mass ratio binary stars ($q \gtrsim 0.6$) in the most reliable clusters \citep[e.g.][]{cordoni_photometric_binaries_2023, donada_multiplicity_fraction_2023a}.

\section{Conclusion}
\label{sec:conclusion}

\paraphrased{In this work, we investigated methods to classify star clusters in the Milky Way as being bound or unbound. By measuring cluster masses and Jacobi radii, we were able to classify 6956 clusters from our catalogue of star clusters in Paper~II as being bound OCs or unbound MGs. This classification method provides a new, more precise way to distinguish between OCs and MGs in \emph{Gaia} data compared to simply using individual cuts on parameters.}


\paraphrased{As a component of this work, we release a catalogue of star cluster masses and radii, which is the largest catalogue of Milky Way cluster masses to date, being around seven times larger than the largest catalogue of OC masses made using \emph{Gaia} data so far \citep{almeida_revisiting_mass_2023}. Our cluster masses were precisely calculated by considering three CMD selection effects and the impact of unresolved binaries. We compare our mass estimates against those in the literature, finding that our masses are typically higher than previous literature results. We suggest that this is due to our inclusion of selection effect corrections.}

\paraphrased{We use our cluster masses to estimate the fraction of clusters from Paper~II that are compatible with bound (instantaneously self-gravitating) objects, publishing an updated star cluster catalogue with improved cluster classifications. Within 15~kpc (the maximum distance that we provide mass measurements for), we find that only 79\% of the clusters from Paper~II are compatible with being bound. Nearby to the Sun, within 250~pc, our catalogue is dominated by MGs, with just 11\% of clusters being compatible with bound objects. Our final catalogue contains 5647 OCs, 3530 of which are in a high-quality sample with higher astrometric S/N and good-quality CMDs. The catalogue contains 1309 MGs, 539 of which are of high quality by the same definition.}

Comparisons between OCs and MGs in our catalogue show interesting differences between these objects. The structural concentration of OCs is a strong function of their mass, and not their age. On the other hand, older MGs are significantly larger than young ones, which is compatible with them being unbound, expanding objects. \referee{Young MGs and OCs in our catalogue appear to form at similar initial sizes, but with MGs expanding significantly more with age.} Our detection of so many MGs in our cluster search in Paper~II was an accident, as we only intended to detect OCs; however, given that both MGs and OCs are remnants of coeval star formation, and some unbound MGs may even be remnants of bound OCs, it would make sense in future blind cluster searches to continue searching for both classes of object and conducting comparisons between them.

We also used these results to derive approximate estimates of the completeness, age function, and mass function of the OC census in \emph{Gaia} DR3. \paraphrased{The completeness of our catalogue is well described by a logarithmic function of only cluster mass up to $\sim$2800~pc, beyond which the 100\% completeness limit does not increase further. \cite{kharchenko_global_2013} stated that the OC census is complete within 1.8~kpc, a claim that has since been disproven by numerous \emph{Gaia}-based works \citep{castro-ginard_new_2018, castro-ginard_hunting_2019, castro-ginard_hunting_open_2020, castro-ginard_hunting_open_2022, liu_catalog_newly_2019,sim_207_2019, hunt_improving_2021, hunt_improving_open_2023}; in this work, we find that our OC census is only approximately 100\% complete at 1.8~kpc for clusters heavier than 230~$\mathrm{M}_\sun$, suggesting that many more low-mass OCs are still yet to be discovered within this distance range.} 

Using this completeness estimate, we confirm the results of \cite{anders_milky_way_2021} that the \emph{Gaia} census of OCs is significantly younger than pre-\emph{Gaia} works. We also derive a completeness-corrected mass function of our OC catalogue, finding that OCs above around $400~\text{M}_\sun$ are compatible with a power law with slope equal to -2, which is compatible with observations of the cluster mass function of numerous other galaxies \citep{krumholz_star_2019}. However, below this mass, we find that there are fewer clusters than expected. Separated into age bins, our cluster mass function appears to flatten with increasing age, suggesting an accelerated rate of cluster dissolution for low-mass clusters. Our cluster mass function implies that the Milky Way should contain a total of around $1.3 \times 10^5$ OCs, $\sim 4\%$ of which are currently known. Finally, in investigation of the mass functions of individual clusters, we find that most OCs are broadly compatible with a Kroupa IMF for ages below 1~Gyr -- but only after extensive correction of their mass functions for selection effects. 

Since the release of \emph{Gaia} DR2 \citep{brown_gaia_2018}, there has been an explosion in studies reporting detections of new OCs \citep[e.g.][]{sim_207_2019,castro-ginard_hunting_open_2020,castro-ginard_hunting_open_2022,liu_catalog_newly_2019,he_catalogue_2021,he_unveiling_hidden_2022,hao_newly_2022}. Works generally agree that an OC should be an overdensity in \emph{Gaia} data, with at least $\sim10$ member stars, and a CMD compatible with a single population of stars. However, until now, there has been no way to further observationally define detected star clusters into bound and unbound objects. The effectiveness of measuring cluster Jacobi radii for this purpose will improve the accuracy and clarity of both the current OC census and future OC censuses based on upcoming data releases.

\begin{acknowledgements}
We thank the anonymous referee for their comments that improved the quality of this paper, as well as Siegfried Röser and Elena Schilbach for further helpful comments. E.L.H. and S.R. gratefully acknowledge funding by the Deutsche Forschungsgemeinschaft (DFG, German Research Foundation) -- Project-ID 138713538 -- SFB 881 (``The Milky Way System'', subproject B5). We thank Josefa Großschedl, Bruno Alessi, Philipp Teutsch, Siegfried Röser, and Elena Schilbach for providing feedback on unreliable clusters or primary names from our Paper~II work. This work has made use of data from the European Space Agency (ESA) mission {\it Gaia} (\url{https://www.cosmos.esa.int/gaia}), processed by the {\it Gaia} Data Processing and Analysis Consortium (DPAC, \url{https://www.cosmos.esa.int/web/gaia/dpac/consortium}). Funding for the DPAC has been provided by national institutions, in particular the institutions participating in the {\it Gaia} Multilateral Agreement. This research has made use of NASA's Astrophysics Data System Bibliographic Services. This research also made use of the SIMBAD database, operated at CDS, Strasbourg, France \citep{wengerm_simbad_astronomical_2000}.

In addition to those cited in the main body of the text, this work made use of the open source Python packages \texttt{NumPy} \citep{harris_array_2020}, \texttt{SciPy} \citep{virtanen_scipy_2020}, \texttt{IPython} \citep{perez_ipython_2007}, \texttt{Jupyter} \citep{kluyver_jupyter_2016}, \texttt{Matplotlib} \citep{hunter_matplotlib_2007}, \texttt{pandas} \citep{mckinney_data_2010, reback2020pandas}, and \texttt{Astropy} \citep{robitaille_astropy_2013, astropycollaboration_astropy_project_2018}. This work also made use of accessible \texttt{Matplotlib}-like colour cycles defined in \cite{petroff_accessible_color_2021}. 

\end{acknowledgements}

\bibliographystyle{aa} 
\bibliography{references} 

\begin{appendix}

\onecolumn
\section{Name updates}

\begin{longtable}{clll}
\caption{\label{tab:catalogue_name_changes}All name updates applied to catalogue.}\\
\hline\hline
ID & Old name & New name & Reason for change \\
\hline
\endfirsthead
\caption{continued.}\\
\hline\hline
ID & Old name & New name & Reason for change \\
\hline
\endhead
\hline
\endfoot
0 & 1636-283 & ESO 452-11 & More common name \citep{perren_unified_cluster_2023} \\
2 & AH03 J0748+26.9 & AH03 J0748-26.9 & Typo \citep{perren_unified_cluster_2023} \\
66 & AT 4 & Alessi-Teutsch 4 & Consistency \\
67 & AT 21 & Alessi-Teutsch 21 & Consistency \\
1316 & Collinder 97 & Teutsch 137 & \cite{kronberger_new_galactic_2006} \\
1706 & FoF 5 & LP 5 & Consistency \\
1707 & FoF 145 & LP 145 & Consistency \\
1708 & FoF 198 & LP 198 & Consistency \\
1709 & FoF 273 & LP 273 & Consistency \\
1710 & FoF 282 & LP 282 & Consistency \\
1711 & FoF 321 & LP 321 & Consistency \\
1712 & FoF 403 & LP 403 & Consistency \\
1713 & FoF 589 & LP 589 & Consistency \\
1714 & FoF 597 & LP 597 & Consistency \\
1715 & FoF 658 & LP 658 & Consistency \\
1716 & FoF 861 & LP 861 & Consistency \\
1717 & FoF 866 & LP 866 & Consistency \\
1718 & FoF 876 & LP 876 & Consistency \\
1719 & FoF 947 & LP 947 & Consistency \\
1720 & FoF 1180 & LP 1180 & Consistency \\
1721 & FoF 1182 & LP 1182 & Consistency \\
1722 & FoF 1218 & LP 1218 & Consistency \\
1723 & FoF 1235 & LP 1235 & Consistency \\
1724 & FoF 1375 & LP 1375 & Consistency \\
1725 & FoF 1428 & LP 1428 & Consistency \\
1726 & FoF 1540 & LP 1540 & Consistency \\
1727 & FoF 1624 & LP 1624 & Consistency \\
1728 & FoF 1800 & LP 1800 & Consistency \\
1729 & FoF 1973 & LP 1973 & Consistency \\
1730 & FoF 1994 & LP 1994 & Consistency \\
1731 & FoF 2068 & LP 2068 & Consistency \\
1732 & FoF 2094 & LP 2094 & Consistency \\
1733 & FoF 2100 & LP 2100 & Consistency \\
1734 & FoF 2117 & LP 2117 & Consistency \\
1735 & FoF 2123 & LP 2123 & Consistency \\
1736 & FoF 2139 & LP 2139 & Consistency \\
1737 & FoF 2143 & LP 2143 & Consistency \\
1738 & FoF 2179 & LP 2179 & Consistency \\
1739 & FoF 2198 & LP 2198 & Consistency \\
1740 & FoF 2210 & LP 2210 & Consistency \\
1741 & FoF 2220 & LP 2220 & Consistency \\
1742 & FoF 2238 & LP 2238 & Consistency \\
1743 & FoF 2253 & LP 2253 & Consistency \\
1744 & FoF 2309 & LP 2309 & Consistency \\
1745 & FoF 2383 & LP 2383 & Consistency \\
1772 & Gulliver 27 & Teutsch 68 & \cite{kronberger_new_galactic_2006} \\
1789 & Gulliver 46 & Teutsch 108 & \cite{kronberger_new_galactic_2006} \\
1893 & HSC 134 & Gran 3 & Is GC \citep{gran_hidden_haystack_2022} \\
2286 & HSC 608 & Teutsch 139 & \cite{kronberger_new_galactic_2006} \\
3201 & HSC 1752 & Teutsch 171 & \cite{kronberger_new_galactic_2006} \\
3693 & HSC 2363 & Teutsch 105 & \cite{kronberger_new_galactic_2006} \\
4111 & HSC 2890 & Gran 4 & Is GC \citep{gran_hidden_haystack_2022} \\
4238 & HXWHB 8 & Teutsch 89 & \cite{kronberger_new_galactic_2006} \\
4323 & Juchert J0644.8+0925 & Juchert J0644.8-0925 & Typo \citep{perren_unified_cluster_2023} \\
4326 & Juchert-Saloran 1 & Juchert-Saloranta 1 & Typo \\
5009 & OC 0431 & Teutsch 62 & \cite{kronberger_new_galactic_2006} \\
5153 & Pal 3 & Palomar 3 & Consistency \\
5154 & Pal 4 & Palomar 4 & Consistency \\
5155 & Pal 13 & Palomar 13 & Consistency \\
5156 & Pal 14 & Palomar 14 & Consistency \\
5379 & TRSG 4 & RSG 4 & Consistency \\
5470 & Teutsch J0718.0+1642 & Teutsch J0718.0-1642 & Typo \citep{perren_unified_cluster_2023} \\
5471 & Teutsch J0924.3+5313 & Teutsch J0924.3-5313 & Typo \citep{perren_unified_cluster_2023} \\
5472 & Teutsch J1037.3+6034 & Teutsch J1037.3-6034 & Typo \citep{perren_unified_cluster_2023} \\
5473 & Teutsch J1209.3+6120 & Teutsch J1209.3-6120 & Typo \citep{perren_unified_cluster_2023} \\
6007 & Theia 5761 & Teutsch 17 & \cite{kronberger_new_galactic_2006} \\
6183 & UBC 212 & Teutsch 59a & \cite{kronberger_new_galactic_2006} \\
6880 & UBC 1564 & Teutsch 86 & \cite{kronberger_new_galactic_2006} \\


\end{longtable}

\end{appendix}

\end{document}